\documentclass[11pt]{article}

\newcommand{\be}{\begin{equation}}
\newcommand{\ee}{\end{equation}}
\usepackage{amsmath}
\usepackage{amssymb}
\usepackage{latexsym}
\usepackage{epsfig}

\usepackage[vcentermath]{youngtab}

\newcommand{\bea}{\begin{eqnarray}}
\newcommand{\eea}{\end{eqnarray}}

\newcommand{\RRR}{{\hbox{\rm R\kern-2.35mm R}}}

\def\ZZZ{{\hbox{ Z\kern-1.6mm Z}}}

\begin{document}

\begin{titlepage}
\rightline{November 2010}
\rightline{\tt MIT-CTP-4193}
\begin{center}
\vskip 2.5cm
{\Large \bf {
Frame-like Geometry of Double Field Theory
}}\\
\vskip 1.0cm
{\large {Olaf Hohm and Seung Ki Kwak}}
\vskip 1cm
{\it {Center for Theoretical Physics}}\\
{\it {Massachusetts Institute of Technology}}\\
{\it {Cambridge, MA 02139, USA}}\\
ohohm@mit.edu, sk$\_$kwak@mit.edu

\vskip 2.5cm
{\bf Abstract}
\end{center}

\vskip 0.5cm

\noindent
\begin{narrower}
We relate two formulations of the recently constructed double field theory to a frame-like
geometrical formalism developed by Siegel. A self-contained presentation of this formalism
is given, including a discussion of the constraints and its solutions, and of the 
resulting Riemann tensor, Ricci tensor and curvature scalar. This curvature
scalar can be used to define an action, and it is shown that this action is equivalent to that
of double field theory.

\end{narrower}

\end{titlepage}

\newpage

\tableofcontents
\baselineskip=16pt

\section{Introduction}
T-duality is one of the simplest and perhaps one of the most intriguing dualities of string theory. It relates the
momentum and winding modes of closed string theory on a torus $T^{D}$ via the non-compact duality group $O(D,D)$ (for a review see \cite{Giveon:1994fu}). It is a fairly natural idea that this duality
symmetry can be made manifest upon introducing `doubled coordinates', both at the level of the world-sheet
\cite{Duff:1989tf,Tseytlin:1990nb,Hull:2004in} and at the level of space-time \cite{Tseytlin:1990nb,Hull:2006va}.
In other words, in addition to the usual coordinates $x^{i}$ associated to momentum modes one has dual coordinates
$\tilde{x}_{i}$ associated to winding modes.

Recently, a formulation of such a `double field theory' has been found
\cite{Hull:2009mi,Hull:2009zb,Hohm:2010jy,Hohm:2010pp} that can be viewed as
an $O(D,D)$ covariantization of the low-energy effective space-time action. 
(For recent papers related to this theory see 
\cite{Berman:2010is,Kwak:2010ew,West:2010ev,Lust:2010iy,Jeon:2010rw}.) 
The conventional action for
the metric $g_{ij}$, the Kalb-Ramond two-form $b_{ij}$ with field strength $H_{ijk}=3\partial_{[i}b_{jk]}$, and the dilaton $\phi$ is given by 
 \bea\label{original}
  S \ = \ \int dx \sqrt{-g}e^{-2\phi}\left[R+4(\partial\phi)^2-\frac{1}{12}H^2\right]\,.
 \eea
The double field theory action is written in terms of the `non-symmetric metric' ${\cal E}_{ij}=g_{ij}+b_{ij}$, which
naturally combines the conventional metric and 2-form, and the dilaton $d$ which is related to
the scalar dilaton $\phi$ via the field redefinition $e^{-2d}=\sqrt{-g}e^{-2\phi}$. It reads
 \bea
\label{THEActionINTRO}
 \begin{split}\hskip-10pt
  S \ = \ \int \,dx d\tilde{x}~
  e^{-2d}\Big[&
  -\frac{1}{4} \,g^{ik}g^{jl}   \,   {\cal D}^{p}{\cal E}_{kl}\,
  {\cal D}_{p}{\cal E}_{ij}
  +\frac{1}{4}g^{kl} \bigl( {\cal D}^{j}{\cal E}_{ik}
  {\cal D}^{i}{\cal E}_{jl}  + \bar{\cal D}^{j}{\cal E}_{ki}\,
  \bar{\cal D}^{i}{\cal E}_{lj} \bigr)~
\\ &    + \bigl( {\cal D}^{i}\hskip-1.5pt d~\bar{\cal D}^{j}{\cal E}_{ij}
 +\bar{{\cal D}}^{i}\hskip-1.5pt d~{\cal D}^{j}{\cal E}_{ji}\bigr)
 +4{\cal D}^{i}\hskip-1.5pt d \,{\cal D}_{i}d ~\Big]\;.
 \end{split}
 \eea
Here, the calligraphic derivatives are defined by
\be
\label{groihffkdf}
{\cal D}_i \ \equiv \ {\partial\over \partial x^i} - {\cal E}_{ik} \,{\partial \over \partial\tilde x_k}\,,
~~~~\bar {\cal D}_i \ \equiv \ {\partial\over \partial x^i} + {\cal E}_{ki}\, {\partial \over \partial\tilde x_k}\,,
\ee
and all indices are raised with  $g^{ij}$, which is the inverse
of the metric $g_{ij} = {1\over 2} ({\cal E}_{ij} + {\cal E}_{ji})$.
As required, the action (\ref{THEActionINTRO}) reduces to (\ref{original}) if the winding derivatives
are set to zero \cite{Hohm:2010jy}. 
Moreover, it is invariant under the T-duality group $O(D,D)$  whose action on the
fields can be written in matrix notation as (generalizing the well-known Buscher rules \cite{Buscher:1987sk})
 \begin{equation}\label{ODDaction}
  {\cal E}^{\prime}(X^{\prime}) \ = \ (a{\cal E}(X)+b)(c{\cal E}(X)+d)^{-1}\;,
  \quad
  d^{\prime}(X^{\prime}) \ = \ d(X)\;, \quad X' = h X\,,
 \end{equation}
 where
  \begin{equation}\label{ODDmatrix}
 h= \begin{pmatrix} a &   b \\ c & d \end{pmatrix} \ \in \ O(D,D)\;,
 \end{equation}
and we have grouped the momentum coordinates $x^{i}$ and the winding coordinates $\tilde{x}_{i}$
into a fundamental $O(D,D)$ vector $X^{M}=(\tilde{x}_{i},x^{i})$.\footnote{We use a notation that does 
not distinguish between compact and non-compact coordinates. In string theory only the compact 
coordinates should be doubled, and in this case the doubling for the non-compact coordinates is 
only formal, with a trivial dependence of the fields on these new coordinates. 
The signature of the duality group $O(D,D)$ applies to the case that the 
space-time metric $g_{ij}$ is positive definite, but all formulas below extend readily to any other 
signature.} 
The action (\ref{THEActionINTRO}) is also invariant under a gauge symmetry with a parameter
$\xi^{M}=(\tilde{\xi}_{i},\xi^{i})$ that combines the conventional diffeomorphism parameter $\xi^{i}$ with
the 1-form gauge parameter $\tilde{\xi}_{i}$ of the 2-form,
 \begin{equation}
 \label{finalgtINTRO}
 \begin{split}
  \delta {\cal E}_{ij} \ &= \ {\cal D}_i\tilde{\xi}_{j}-\bar{{\cal D}}_{j}\tilde{\xi}_{i}
  +\xi^{M}\partial_{M}{\cal E}_{ij}
  +{\cal D}_{i}\xi^{k}{\cal E}_{kj}+\bar{\cal D}_{j}\xi^{k}{\cal E}_{ik}\;,\\[0.5ex]
 \delta d \ &= \ \xi^M \partial_M d - {1\over 2}  \partial_M \xi^M\,,
 \end{split}
 \end{equation}
and which reduces to the familiar diffeomorphism and 2-form gauge symmetry for $\tilde{\partial}=0$.

The consistency of the action (\ref{THEActionINTRO}) requires the constraint that all fields and gauge
parameters and all their products are annihilated by the differential operator $\tilde{\partial}^{i}\partial_{i}$.
This is a stronger version of the level-matching condition of closed string theory, 
and will therefore sometimes be referred to as `the strong constraint'. 
It takes a manifestly $O(D,D)$ covariant form, upon introducing the $O(D,D)$ invariant
metric $\eta$:   We require
 \be\label{ODDconstr}
   \partial^{M}\partial_{M}A \ = \ \eta^{MN}\partial_{M}\partial_{N}A \ = \ 0\;, \qquad
   \partial^{M}A\,\partial_{M}B \ = \ 0\;, \qquad
   \eta^{MN} =  \begin{pmatrix}
    0&1 \\1&0 \end{pmatrix}\,,
 \ee
for all fields and parameters $A,B$. This constraint implies that locally there is always an $O(D,D)$ transformation
that rotates into a `T-duality frame' where the fields depend only on half of the coordinates, for instance  the momentum coordinates.

Despite taking a strikingly simple form, the properties of the double field theory action (\ref{THEActionINTRO})
are not very transparent. Even though the $O(D,D)$ invariance is well understood (and `manifest' in the
sense that each term is separately invariant \cite{Hohm:2010jy}), the fields do not transform in linear representations of $O(D,D)$. Moreover, the gauge symmetry is far from being manifest
in the formulation (\ref{THEActionINTRO}), and therefore a more geometrical understanding is desirable.

More recently, a reformulation of (\ref{THEActionINTRO}) has been given in which some of these features are
more accessible \cite{Hohm:2010pp}. It is based on the `generalized metric'
 \be\label{genmetric}
  {\cal H}^{MN} \ = \  \begin{pmatrix}    g_{ij}-b_{ik}g^{kl}b_{lj} & b_{ik}g^{kj}\\[0.5ex]
  -g^{ik}b_{kj} & g^{ij}\end{pmatrix}\;,
 \ee
which combines the metric and 2-form in such a way that it transforms covariantly under $O(D,D)$ according to its index structure, i.e., in a linear representation 
as opposed to the non-linear representation of ${\cal E}_{ij}$ above. The double field theory
action can then be written in the manifestly $O(D,D)$ invariant form
 \bea\label{Hactionx}
 \begin{split}
  S \ = \ \int dx d\tilde{x}\,e^{-2d}~\Big(~&\frac{1}{8}\,{\cal H}^{MN}\partial_{M}{\cal H}^{KL}
  \,\partial_{N}{\cal H}_{KL}-\frac{1}{2}{\cal H}^{MN}\partial_{N}{\cal H}^{KL}\,\partial_{L}
  {\cal H}_{MK}\\
  &-2\,\partial_{M}d\,\partial_{N}{\cal H}^{MN}+4{\cal H}^{MN}\,\partial_{M}d\,
  \partial_{N}d~\Big)\,.
 \end{split}
 \eea
Here, the derivatives $\partial_{M}=(\tilde{\partial}^{i},\partial_{i})$ and
$\partial^{M}=\eta^{MN}\partial_{N}=(\partial_{i},\tilde{\partial}^{i})$ transform covariantly under
$O(D,D)$.
Remarkably, in terms of ${\cal H}^{MN}$ the gauge symmetry parameterized by $\xi^{M}$ becomes manifestly $O(D,D)$ invariant,
 \be\label{manifestH}
  \delta_{\xi}{\cal H}^{MN} \ = \ \xi^{P}\partial_{P}{\cal H}^{MN}
  +(\partial^{M}\xi_{P} -\partial_{P}\xi^{M})\,{\cal H}^{PN}
  +
 ( \partial^{N}\xi_{P} -\partial_{P}\xi^{N})\,{\cal H}^{MP}\;.
 \ee
Here and in the following $O(D,D)$ indices are raised and lowered with $\eta^{MN}$.
This form of the gauge transformations naturally suggests a notion of `generalized Lie derivative',
in which each index gives rise to a covariant and a contravariant contribution. In the formulation 
(\ref{Hactionx}) the gauge
invariance of the double field theory action can be checked more easily, although it is still non-manifest.

For both formulations presented above the action can actually be written in an Einstein-Hilbert like form
with a scalar curvature ${\cal R} ={\cal R}({\cal E},d)={\cal R}({\cal H},d)$ that can be viewed 
as a function of $d$ and either ${\cal E}_{ij}$ or ${\cal H}^{MN}$.
More precisely, up to boundary terms, (\ref{THEActionINTRO})
and (\ref{Hactionx}) can be written as
 \bea\label{masteractionINTRO}
  S \ = \ \int dxd\tilde{x}\,e^{-2d}\,{\cal R}({\cal E},d) \ = \  \int dxd\tilde{x}\,e^{-2d}\,{\cal R}({\cal H},d)\;.
 \eea
In here, ${\cal R}$ transforms as a scalar and $e^{-2d}$ as a density,
 \be\label{covtrans}
   \delta_{\xi}{\cal R} \ = \ \xi^{M}\partial_{M}{\cal R}\;, \qquad
   \delta_{\xi}\big(e^{-2d}\big) \ = \ \partial_{M}\big(\xi^{M}e^{-2d}\big)\;,
 \ee
from which invariance of the action immediately follows. The scalar curvature has, however, only been
determined `by hand' as functions of $d$ and ${\cal E}_{ij}$ or ${\cal H}^{MN}$, respectively, by requiring
the transformation behavior (\ref{covtrans}). Again, a more geometrical understanding, in which
${\cal R}$ arises from a Riemann-tensor-like object that is manifestly covariant, is desirable.

\vspace{0.15cm}

Prior to these developments Siegel has
introduced some time ago a duality-covariant
geometrical formalism in a remarkable paper \cite{Siegel:1993th} (extending the results of
\cite{Siegel:1993xq}).
This formalism is based on a frame-field $e_{A}{}^{M}$ that carries a flat index $A$
corresponding to a
local $GL(D)\times GL(D)$ symmetry. 
This direct product structure with two independent general linear groups reflects the left-right factorization of 
closed string theory. 
The formalism features connections for this
local symmetry and
covariant curvature tensors. Intriguingly, it has the same transformations under
$\xi^{M}$ according
to the generalized Lie derivatives  as in (\ref{manifestH}), and it requires the
same constraint (\ref{ODDconstr}). Given these and other similarities it is natural to assume that, upon suitable
identifications and gauge fixings, the
formalism of Siegel is
in fact equivalent to the double field theory formulation reviewed above.
In this paper we will show that this is indeed the case.

In \cite{Hohm:2010pp} this relation has already been elaborated at the level of
the field content
and the symmetry transformations. Here, we go beyond that by relating the
curvature scalars
appearing in (\ref{masteractionINTRO}) to the curvature tensor for the
$GL(D)\times GL(D)$ connections
of Siegel's formulation. In doing so we believe to both clarify the geometrical
meaning of the recent results on double field theory and to give a more explicit and thereby more
accessible treatment of Siegel's
formalism.

This paper is organized as follows. In sec.~2 we review the $O(D,D)$ covariant generalized Lie derivatives
and Siegel's frame-like geometrical formalism. In sec.~3 we discuss the general action principle and 
derive the Bianchi identities implied by gauge invariance. These two sections are mainly a review 
of \cite{Siegel:1993th} and \cite{Hohm:2010pp}, 
but also contain novel results, as the manifestly $O(D,D)$ and 
$GL(D)\times GL(D)$ covariant form (\ref{explR}) of the scalar curvature. 
The main results of this paper are given in sec.~4 and 5, where we relate the frame formalism 
to the explicit formulations in terms of ${\cal E}_{ij}$ and ${\cal H}^{MN}$. 
Specifically, in sec.~4 we show the equivalence of the scalar curvature and the corresponding scalar 
found in \cite{Hohm:2010jy}, and relate 
in particular `$O(D,D)$ covariant derivatives' introduced there to the $GL(D)\times GL(D)$
connections. In sec.~5 we give an independent proof of the equivalence of the curvature scalars
in the formulation with ${\cal H}^{MN}$ given in \cite{Hohm:2010pp}.
We close with a summary and outlook in sec.~6. Some technically involved calculations related
to the curvature tensor can be found in the appendix.

\section{Geometrical frame formalism}\setcounter{equation}{0}
In this section we first review the
novel gauge transformations parametrized by $\xi^{M}$ and the associated C-bracket. 
Next we introduce frame fields which are subject to the tangent space symmetry $GL(D)\times GL(D)$
together with connections for this symmetry. Finally, a covariant curvature tensor is discussed.

\subsection{Generalized Lie derivatives, Courant bracket and frame fields}
The generalized Lie derivative is defined for tensors with an arbitrary number of upper and lower
$O(D,D)$ indices by the straightforward extension of
 \be
 \label{genLie}
  \widehat{{\cal L}}_{\xi} A_{M}{}^{N} \ \equiv \
  \xi^{P}\partial_{P}A_{M}{}^{N}
+ (\partial_{M}\xi^{P}-\partial^{P}\xi_{M})
  \,A_{P}{}^{N}
  +(\partial^{N}\xi_{P}\,    -\partial_{P}\xi^{N})       A_{M}{}^{P}\,.
 \ee
With this definition the gauge transformation (\ref{manifestH}) simply reduces to
\be
 \delta_{\xi}{\cal H}^{MN} \ = \ \widehat{{\cal L}}_{\xi}{\cal H}^{MN}\,.
\ee
In general we will refer to $O(D,D)$ tensors that transform according to the generalized
Lie derivative under gauge transformations parameterized by $\xi^{M}$ as `generalized tensors' or as transforming covariantly under $\xi^{M}$.

An important consistency property of this formalism is that the $O(D,D)$ invariant metric
that is used in (\ref{genLie}) to raise and lower indices has vanishing generalized
Lie derivative,
 \be
   \widehat{\cal L}_{\xi}\eta^{MN} \ = \ \xi^{P}\partial_{P}\,\eta^{MN}-\partial^{N}\xi^{M}
  -\partial^{M}\xi^{N}+\partial^{N}\xi^{M}+\partial^{M}\xi^{N} \ = \ 0\;.
 \ee
Accordingly, in this formalism it is consistent to have this \textit{constant} tensor with two upper
or two lower `curved' or `world' indices.

The closure of the gauge transformations spanned by $\xi^{M}$ or, equivalently, the algebra
of generalized Lie derivatives can be straightforwardly determined in this formulation and is
governed by the `C-bracket',
 \be
   \bigl[ \,  \widehat{{\cal L}}_{\xi_1}\,,  \widehat{{\cal L}}_{\xi_2} \, \bigr]\, 
   \ = \ - \widehat{{\cal L}}_{[\xi_1, \xi_2]_{{}_{\rm C}}}   \,,
 \ee
where
\be
 \label{cbracketdef}
  \bigl[ \xi_1,\xi_2\bigr]_{\rm{C}}^{M} \ \equiv
   \ \xi_{1}^{N}\partial_{N}\xi_{2}^M -\frac{1}{2}\,  \xi_{1}^P\partial^{M}\xi_{2\,P}
   -(1\leftrightarrow 2)\;.
 \ee
This bracket is the $O(D,D)$ covariant double field theory extension of the Courant bracket of
generalized geometry \cite{Tcourant,Hitchin,Gualtieri}, as has been shown in \cite{Hull:2009zb}.
An important property that will be used later is that the C-bracket of two generalized vectors
is again a generalized vector.
In order to verify this let $X^{M}$ and $Y^{M}$ be
transforming as $\delta_{\xi}X^{M}=\widehat{\cal L}_{\xi}X^{M}$ and
$\delta_{\xi}Y^{M}=\widehat{\cal L}_{\xi}Y^{M}$, respectively.
For the computation of the gauge variation of their C-bracket it is useful to keep in mind that the variation of an $O(D,D)$ invariant
expression automatically combines into the covariant terms according to the generalized Lie derivative and into non-covariant terms that originate exclusively from the variation of partial derivatives. Thus, we find
 \begin{eqnarray}
  \delta_{\xi}\bigl[ X,Y\bigr]_{\rm C}^{M} &=&
  \delta_{\xi}\Big(X^{N}\partial_{N}Y^{M}-\frac{1}{2}X^{P}\partial^{M}Y_{P}-(X\leftrightarrow Y)\Big) \\ \nonumber
  &=& \widehat{\cal L}_{\xi}\bigl[ X,Y\bigr]_{\rm C}^{M}\\ \nonumber
  &&+X^{N}\partial_{N}(\partial^{M}\xi_{K}-\partial_{K}\xi^{M})Y^{K}
  -\frac{1}{2}X^{P}\partial^{M}(\partial_{P}\xi^{K}
  -\partial^{K}\xi_{P})Y_{K}-(X\leftrightarrow Y) \\ \nonumber
  &=&  \widehat{\cal L}_{\xi}\bigl[ X,Y\bigr]_{\rm C}^{M} \;,
\end{eqnarray}
where the
cancelation of the non-covariant terms easily follows from the antisymmetry in $X$ and $Y$.
This establishes the covariance of the C-bracket. As the variation in the first line can also be written as
$[\delta X,Y]_{\rm C}+[X,\delta Y]_{\rm C}$, this covariance property of the C-bracket can be
put more compactly as
 \be
  \widehat{\cal L}_{\xi}\bigl[ X,Y\bigr]_{\rm C}
  \ = \ \big[\widehat{\cal L}_{\xi}X,Y\big]_{\rm C}+\big[X,\widehat{\cal L}_{\xi}Y\big]_{\rm C}\;,
 \ee
which is the analogue of the invariance of the Lie bracket under the usual Lie derivative.

In general, partial derivatives of generalized tensors are not generalized tensors. An exception
is a generalized scalar $S$ which according to (\ref{genLie}) simply transforms as
 \be
  \delta_{\xi}S \ = \ \widehat{\cal L}_\xi  S \ = \ \xi^P \partial_P S \,.
 \ee
Therefore, its partial derivative transforms as
 \be
  \delta_{\xi}(\partial_{M}S) \ = \ \partial_{M}(\xi^{P}\partial_{P}S) \ = \
  \xi^{P}\partial_{P}(\partial_{M}S)+(\partial_{M}\xi^{P}-\partial^{P}\xi_{M})\partial_{P}S
  \ \equiv \ \widehat{\cal L}_{\xi}(\partial_{M}S)\;,
 \ee
where in the second equality we were allowed to add the third term because it is zero by the
constraint  (\ref{ODDconstr}).  Thus, $\partial_{M}S$ transforms covariantly, i.e., as a generalized
covariant tensor. This covariant transformation behavior does not hold for partial derivatives of higher
tensors, not even for antisymmetrized combinations like $\partial_{[M}V_{N]}$ --- in contrast to conventional diffeomorphisms.

In the following we will introduce a frame field which allows to convert arbitrary tensors from
`world'-tensors into `tangent space'-tensors and thereby into scalars under $\xi^{M}$.
Specifically, following Siegel \cite{Siegel:1993th} we introduce a frame field $e_{A}{}^{M}$,
which is a generalized vector and has a flat index $A$ corresponding to a local
$GL(D)\times GL(D)$ symmetry, i.e.,\footnote{We note that our conventions for the frame field
differ from those in \cite{Hohm:2010pp} (c.f.~eq.~(5.12)) in order to be more in line with \cite{Siegel:1993th}.}
 \be\label{Egauge}
   e_{A}{}^{M} \ = \ \begin{pmatrix} e_{ai} &  e_{a}{}^{i} \\ e_{\bar{a}i} & e_{\bar{a}}{}^{i} \end{pmatrix}
   \;.
 \ee
We assume this vielbein to be invertible and denote the inverse by $e_{M}{}^{A}$.  
In (\ref{Egauge}) we used the splitting $M = (\,{}_{i}\,,\,{}^{i}\,)$ of the $O(D,D)$ index and $A=(a,\bar{a})$
is the $GL(D)\times GL(D)$ index. Given the $O(D,D)$ invariant metric $\eta_{MN}$ we can build
an $X$-dependent `tangent space' metric of signature $(D,D)$,
 \be\label{flatmetric}
  {\cal G}_{AB} \ = \ e_{A}{}^{M}\,e_{B}{}^{N}\,\eta_{MN}\;,
 \ee
with inverse ${\cal G}^{AB}=\eta^{MN}e_{M}{}^{A}e_{N}{}^{B}$, 
which will be used to raise and lower flat indices. 
The raising and lowering of world indices with $\eta$ and of flat indices with ${\cal G}$ is consistent 
with inverting the frame field (\ref{Egauge}) in that 
 \be
  e_{M}{}^{A} \ = \ \eta_{MN}{\cal G}^{AB}e_{B}{}^{N} \qquad\Rightarrow\qquad
  e_{M}{}^{A}e_{A}{}^{N} \ = \ \delta_{M}{}^{N}\;,
 \ee
as follows from the definition (\ref{flatmetric}).   
In order for $e_{A}{}^{M}$ to describe only the physical degrees of freedom it turns out to be necessary 
to impose the $GL(D)\times GL(D)$ covariant constraint
 \be\label{Goff}
   {\cal G}_{a\bar{b}} \ = \ 0 \quad \Leftrightarrow \quad e_{(a}{}^{i}\,e_{\bar{b})i} \ = \ 0\;,
 \ee
which is related to the left-right factorization of closed string theory \cite{Siegel:1993th}.\footnote{An alternative 
motivation of this constraint starting from
generalized geometry and the generalized metric ${\cal H}$ has been given in \cite{Hohm:2010pp},
c.f.~the discussion after eq.~(\ref{Hroot}) below.}

Using the frame field one can introduce a `flattened' derivative $e_{A}$, defined by 
 \be\label{flatder}
  e_{A} \ \equiv \ e_{A}{}^{M}\,\partial_{M}\;.
\ee
We note that the strong constraint (\ref{ODDconstr}) takes the following form in terms of flat indices,
 \be\label{flatconstr}
  e^{A}X\,e_{A}Y \ = \ {\cal G}^{AB}e_{A}{}^{M}e_{B}{}^{N}\partial_{M}X\,\partial_{N}Y
  \ = \ \eta^{MN}\partial_{M}X\,\partial_{N}Y \ = \  0\;,
 \ee
for arbitrary functions $X$ and $Y$.
Due to the covariance of the partial derivative of a generalized scalar discussed above,
the action of $e_{A}$ on an arbitrary tensor with only flat indices, $e_{A}X_{BC\ldots}$, is covariant
under $\xi^{M}$ transformations. Of course, it will not be covariant under the local frame rotations, and
so covariant derivatives have to be introduced.  Thereby, the problem of defining derivative
operations that are covariant under generalized diffeomorphisms parameterized by $\xi^{M}$ has been
converted to the problem of introducing covariant derivatives and connections for the $GL(D)\times GL(D)$
tangent space symmetry, to which we turn now.

\subsection{$GL(D)\times GL(D)$ connections and constraints}
We define the infinitesimal local $GL(D)\times GL(D)$ transformations to be
 \be\label{GLtrans}
  \delta_{\Lambda}V_{A} \ = \ \Lambda_{A}{}^{B}V_{B}\;, \qquad
  \delta_{\Lambda}V^{A} \ = \ -\Lambda_{B}{}^{A}V^{B}\;,
 \ee
and analogously for tensors with an arbitrary number of upper and lower indices. Since 
we are dealing with $GL(D)\times GL(D)$, the non-vanishing parameters are
$\Lambda_{a}{}^{b}$ and $\Lambda_{\bar{a}}{}^{\bar{b}}$. Covariant
derivatives with flattened indices are given by
 \be\label{covder}
  \nabla_{A}V_B \ = \ e_{A}V_{B}+\omega_{AB}{}^{C}V_{C}\;, \qquad
  \nabla_{A}V^{B} \ = \ e_{A}V^{B}-\omega_{AC}{}^{B}V^{C}\;,
 \ee
where we have introduced connections $\omega_{AB}{}^{C}$. Again, since we are dealing with
gauge group $GL(D)\times GL(D)$ the only non-vanishing components of the connections are
 \be\label{nonzero}
   \omega_{AB}{}^{C}\;:\qquad \omega_{Ab}{}^{c}\;, \quad \omega_{A\bar{b}}{}^{\bar{c}}\;.
 \ee
Moreover, the constraint (\ref{Goff}) implies that the same holds for connections 
with all indices lowered. 
We will frequently make use of the fact that components like $\omega_{ab}{}^{\bar{c}}$ 
and $\omega_{ab\bar{c}}$ vanish.
We require that the connections transform under $\xi^{M}$ as scalars and therefore, as
discussed above, the covariant derivatives (\ref{covder}) transform as scalars, too. They transform
also covariantly under $GL(D)\times GL(D)$ if we require that the $\omega_{AB}{}^{C}$ transform
as connections, i.e.,
 \be\label{connvar}
  \delta \omega_{Aa}{}^{b} \ = \ -\nabla_{A}\Lambda_{a}{}^{b} + \Lambda_{A}{}^{B}
  \omega_{Ba}{}^{b}\;, \qquad
  \nabla_{A}\Lambda_{a}{}^{b} \ = \ e_{A}\Lambda_{a}{}^{b}+\omega_{Aa}{}^{c}
  \Lambda_{c}{}^{b}-\omega_{Ac}{}^{b}\Lambda_{a}{}^{c}\;,
 \ee
and analogously for barred indices. We note that the additional term in $\delta \omega_{Aa}{}^{b}$
as compared to the familiar transformation rule for a Yang-Mills gauge potential is due to the
conversion of the 1-form index into a flat one.

Next we have to impose covariant constraints that allow us to solve for (part of) the connections in terms of the physical fields. A natural starting point is the C-bracket governing the gauge algebra. In ordinary Riemannian geometry the torsion constraint of the Levi-Civita connection implies that in the
Lie bracket of two vector fields the partial derivatives can be replaced by covariant derivatives.
In the double field theory context the Lie bracket is replaced by the C-bracket in that only the latter
transforms covariantly under generalized diffeomorphisms.  Since we are dealing here with
flattened derivatives it is thus natural to define a torsion tensor in such a way that it vanishes if and only if
in the C-bracket with flattened parameter
$\xi_{12}^{A} =  \big[\xi_{1},\xi_{2}\big]_{\rm C}^{M}e_{M}{}^{A}$
the partial derivatives are replaced by $GL(D)\times GL(D)$ covariant derivatives, i.e.,
 \be\label{Storsion}
  \xi_{12}^{A}  \ = \
  \xi_1^{B}\nabla_{B}\xi_2^{A}-\frac{1}{2}\xi_{1B}\nabla^{A}\xi_{2}^{B}
  -   (1\leftrightarrow 2) +\xi_1^{B}\xi_2^{C} \,{\cal T}_{BC}{}^{A}\;.
 \ee
This leads to\footnote{In this paper we employ the convention that symmetrization and anti-symmetrization involves the combinatorial factor, e.g., $X_{[ab]}=\tfrac{1}{2}(X_{ab}-X_{ba})$. In some formulas this leads to numerical factors  that are different from those in \cite{Siegel:1993th}.}
 \be\label{Storsionsol}
  {\cal T}_{AB}{}^{C} \ = \ \Omega_{AB}{}^{C}
  +2\left(\omega_{[AB]}{}^{C}+\frac{1}{2}\omega^{C}{}_{[AB]}\right)\;,
 \ee
where
 \be\label{newanhol}
  \Omega_{AB}{}^{C} \ = \ 2\left(f_{[AB]}{}^{C}+\frac{1}{2}f^{C}{}_{[AB]}\right)\;, \qquad
  f_{ABC} \ \equiv \ (e_{A}e_{B}{}^{M})e_{CM}\;.
 \ee
Another more covariant form of the torsion tensor is
 \be\label{GLcovT}
  {\cal T}_{AB}{}^{C}
   \ = \ 2\left(e_{N}{}^{C}\nabla_{[A}e_{B]}{}^{N}-\frac{1}{2}e_{N[A}\nabla^{C}e_{B]}{}^{N}\right)\;.
 \ee
We note here that the torsion tensor defined like this does not coincide with the usual definition via
the commutator of covariant derivatives. We will return to this issue below.

The $\Omega_{AB}{}^{C}$ introduced above can be seen as generalized `coefficients of anholonomy'
in that
 \be\label{anhol}
  \big[ e_{A},e_{B} \big] \ = \ \Omega_{AB}{}^{C}e_{C}\;.
 \ee
To be more precise, the first term in (\ref{newanhol}) proportional to $f_{[AB]}{}^{C}$ corresponds to the usual coefficients of anholonomy,
while the second term drops out in the equation (\ref{anhol}) by virtue of the constraint (\ref{flatconstr}).
The full $\Omega_{AB}{}^{C}$ are obtained unambiguously from the C-bracket,
 \be\label{Canholonomy}
  \big[ e_{A},e_{B}\big]_{\rm C}^{M} \ = \ \Omega_{AB}{}^{C}\,e_{C}{}^{M}\;.
 \ee
This can be verified by inserting the components $e_{A}{}^{M}$ into the definition (\ref{cbracketdef}). 
These generalized coefficients of anholonomy,
as opposed to the usual ones and the $f_{ABC}$, are fully covariant under $\xi^{M}$ transformations.
This follows directly from (\ref{Canholonomy}) and the fact that the C-bracket transforms covariantly.
Since the $\omega_{AB}{}^{C}$ are generalized scalars it follows from (\ref{Storsionsol})
that the torsion tensor transforms covariantly under $\xi^{M}$, while its covariance under
$GL(D)\times GL(D)$ is manifest in the form (\ref{GLcovT}). Alternatively, this covariance can
be inferred from the defining equation (\ref{Storsion}) and the covariance of the C-bracket. In total,
imposing the torsion constraint
 \be\label{TORsion}
  {\cal T}_{AB}{}^{C} \ = \ 0
 \ee
is consistent with all symmetries.

Next, we impose the `metricity condition' that the metric ${\cal G}_{AB}$ is covariantly constant  \cite{Siegel:1993th},
 \be\label{covconst}
  \nabla_{A}{\cal G}_{BC} \ = \ 0 \qquad \Leftrightarrow \qquad
  e_{A}{\cal G}_{BC}+2\omega_{A(BC)} \ = \ 0
  \;.
 \ee
We recall that indices (here on $\omega$)
are lowered with ${\cal G}_{AB}$.

Finally, we impose a constraint that allows for partial integrations in an
action using the covariant derivatives  \cite{Siegel:1993th}. Specifically, as actions are defined with the density $e^{-2d}$, we require
 \be\label{dilconstr}
  \int e^{-2d}\, V\nabla_{A}V^{A} \ = \ -\int e^{-2d}\,V^{A}\nabla_{A}V \ = \
  -\int e^{-2d}\,V^{A}e_{A}V\;
 \ee
for arbitrary $V$ and $V^{A}$.  The consistency of this and the previous constraints will be
confirmed in the next subsection by providing the explicit solutions.

\subsection{Solving the constraints}
We solve now the above constraints and show their mutual consistency. First, the metricity condition
(\ref{covconst}) can be trivially solved,
 \be\label{metrsol}
  \omega_{A(BC)} \ = \ -\frac{1}{2}e_{A}{\cal G}_{BC} \ = \ -f_{A(BC)}\;,
 \ee
and determines the part symmetric in the `group indices' of $\omega_{ABC}$ completely.

We turn next to the torsion constraint (\ref{TORsion}), which reads explicitly
 \be\label{constr}
  \Omega_{ABC} \ = \ \omega_{BAC}+\frac{1}{2}\omega_{CBA}-\omega_{ABC}
  -\frac{1}{2}\omega_{CAB}  \;.
 \ee
To simplify this further we decompose $\omega_{ABC}$ into a part which is antisymmetric
in its last two indices and into a part which is symmetric in its last two indices,
 \begin{eqnarray}\label{torsionform}\nonumber
  \Omega_{ABC} &=& \omega_{B[AC]}+\omega_{B(AC)}
  +\frac{1}{2}\big(\omega_{C[BA]}+\omega_{C(BA)}\big)
  -\omega_{A[BC]}-\omega_{A(BC)}-\frac{1}{2}\big(\omega_{C[AB]}
  +\omega_{C(AB)}\big)
   \\
  &=& 3\omega_{[BAC]}  +\omega_ {B(AC)}-\omega_{A(BC)}
    \;.
 \end{eqnarray}
 It follows that the completely antisymmetric part is fully determined,
 \be\label{antisol}
  \omega_{[ABC]} \ = \ -\frac{1}{3}\Omega_{[ABC]} \ = \ -f_{[ABC]} \;.
 \ee
In order to gain further insights from (\ref{torsionform}) we decompose the $GL(D)\times GL(D)$
indices and use that only $\omega_{Aab}$ and $\omega_{A\bar{a}\bar{b}}$ are non-zero.
This leads to
 \be
  \Omega_{a\bar{b}\bar{c}} \ = \ 3\omega_{[\bar{b}a\bar{c}]}
  +\omega_{\bar{b}(a\bar{c})}-\omega_{a(\bar{b}\bar{c})} \ = \ -\omega_{a[\bar{b}\bar{c}]}
  -\omega_{a(\bar{b}\bar{c})} \ = \ -\omega_{a\bar{b}\bar{c}}\;,
 \ee
and similarly for the opposite index structure. Thus, in total
 \be\label{Omcomp}
  \omega_{a\bar{b}\bar{c}} \ = \ -\Omega_{a\bar{b}\bar{c}}\;, \qquad
  \omega_{\bar{a}bc} \ = \ -\Omega_{\bar{a}bc}\;.
 \ee
These equations determine, in particular, symmetric parts as $\omega_{a(\bar{b}\bar{c})}$ which were
already given by (\ref{metrsol}). They are, however, consistent as can be confirmed by an explicit
computation using (\ref{newanhol}),
 \be
   \omega_{a(\bar{b}\bar{c})} \ = \ -\Omega_{a(\bar{b}\bar{c})} \ = \ -\frac{1}{2}e^{}_{a}{\cal G}_{\bar{b}\bar{c}}\;,
 \ee
 and similarly for the opposite index structure.

Finally, we solve the constraint (\ref{dilconstr}), which after integration by parts reads explicitly
  \begin{eqnarray}
  \int e^{-2d}\, V\nabla_{A}V^{A} &=&
  \int e^{-2d}\, V\left(e_{A}{}^{M}\partial_{M}V^{A}-\omega_{AB}{}^{A}V^{B}\right)
  \\ \nonumber
  &=&  -\int e^{-2d}\,V^{A}\left(e_{A}{}^{M}\partial_{M}V
  +e^{2d}V\partial_{M}\big(e_{A}{}^{M}e^{-2d}\big)
  +V\omega_{BA}{}^{B} \right)\;.
 \end{eqnarray}
From this we read off
 \be\label{tracepart}
  \omega_{BA}{}^{B} \ = \ -\tilde{\Omega}_{A} \ \equiv \ -e^{2d}\partial_{M}\big(e_{A}{}^{M}e^{-2d}\big)
  \ = \ -\partial_{M}e_{A}{}^{M}+2e_{A}d\;,
 \ee
where we introduced $\tilde{\Omega}_{A}$ for notational convenience. This constraint can be interpreted
as setting the
following `new torsion' to zero,
\be\label{Ttilde}
  \tilde{T}_{A} \ = \ \partial_{M}e_{A}{}^{M}+\omega_{BA}{}^{B}-2e_A d \ = \  0\;,
\ee
which yields for a scalar $S$
 \be\label{Ttilde2}
   \nabla_A \nabla^A S \ = \ - \tilde{T}_A \nabla^A S \ = \ 0 \, ,
 \ee
where the first equality follows by virtue of the strong constraint (\ref{flatconstr}).

We conclude this section by summarizing which connections are determined by the above 
constraints (\ref{TORsion}), (\ref{covconst}) and (\ref{dilconstr}). 
First, the `off-diagonal' components $\omega_{\bar{a}bc}$ and $\omega_{a\bar{b}\bar{c}}$
are completely determined according to (\ref{Omcomp}). 
For the `diagonal' components $\omega_{abc}$
and $\omega_{\bar{a}\bar{b}\bar{c}}$ the parts symmetric in the last two indices are fully 
determined by (\ref{metrsol}). Therefore, it is sufficient for the remaining components to 
focus on the part antisymmetric in the last two indices, whose irreducible parts, say for $\omega_{abc}$, 
are given by the following tensor product
 \be\label{tensorpr}
  \omega_{a[bc]} \;: \qquad
  {\small \yng(1)}\;\otimes \;{\small \yng(1,1)} \ \; = \; \ 
  {\small \yng(1,1,1)}\;\oplus\; {\small \yng(2,1)} \; ,
 \ee 
where the Young tableaux refer to the left $GL(D)$ group. In here, 
the completely antisymmetric part $\omega_{[abc]}$ is determined by
(\ref{antisol}). For the `mixed-Young tableaux' representation on the right-hand side
of (\ref{tensorpr}) the trace parts are determined by (\ref{tracepart}) in terms of the dilaton,
leaving precisely the trace-free part of this $(2,1)$ representation as the undetermined connections.
Its dimension is given by $\tfrac{1}{3}D(D+2)(D-2)$ and therefore, taking the right 
$GL(D)$ into account,  the number of undetermined components is twice this value. 
That not all components are determined by the above constraints limits the extent to which invariant curvatures can be constructed out of the physical fields, which will be discussed in the next
subsection.

\subsection{Covariant cuvature tensor}
Let us now turn to the construction of invariant curvature tensors for the $GL(D)\times GL(D)$ connections.
In general, given covariant derivatives one can define curvatures through their commutator, say, acting
on $V_{C}$,
 \be\label{gencomm}
  \big[ \nabla_{A},\nabla_{B}\big] V_{C} \ = \
  T_{AB}{}^{D}\nabla_{D}V_{C}
  +R_{ABC}{}^{D}V_{D}\;.
 \ee
This leads to the standard expressions
 \begin{eqnarray}\label{naivecur}
  T_{AB}{}^{C} &=& \Omega_{AB}{}^{C}+2\omega_{[AB]}{}^{C}\;, \\ \label{naiveR}
  R_{ABC}{}^{D} &=& e_{A}\omega_{BC}{}^{D}-e_{B}\omega_{AC}{}^{D}
  +\omega_{AC}{}^{E}\omega_{BE}{}^{D}-\omega_{BC}{}^{E}\omega_{AE}{}^{D}
  -\Omega_{AB}{}^{E}\omega_{EC}{}^{D}\;.
 \end{eqnarray}
We note that the torsion tensor $T_{AB}{}^{C}$ defined like this does not coincide with the torsion
${\cal T}_{AB}{}^{C}$ defined earlier in (\ref{Storsionsol}). 
Given the modification of the $\xi^{M}$ gauge transformations as compared to the standard 
diffeomorphisms it was, however, only consistent to 
set ${\cal T}_{AB}{}^{C}=0$. 
We conclude that the \textit{conventional} torsion is necessarily non-zero when imposing
(\ref{TORsion}). More precisely, comparing (\ref{Storsionsol}) with (\ref{naivecur}) one finds
the non-vanishing torsion 
 \be
  T_{AB}{}^{C} \ = \ -\omega^{C}{}_{[AB]}\;.
 \ee
Consequently, the commutator (\ref{gencomm}) of covariant derivatives
reads
 \be\label{finalcomm}
  \big[ \nabla_{A},\nabla_{B}\big] V_{C} \ = \
  -\omega^{D}{}_{[AB]}\nabla_{D}V_{C}
  +R_{ABC}{}^{D}V_{D}\;.
 \ee
An immediate consequence is that $R_{ABC}{}^{D}$ as defined in (\ref{naiveR}) cannot be fully
covariant with respect to $GL(D)\times GL(D)$, because the left-hand side of (\ref{finalcomm}) is manifestly covariant but the right-hand side contains a bare gauge field.

At this stage a comment is in order regarding the non-covariance of the curvature tensor $R$,
because formally it coincides with a conventional field strength (with flattened indices) that
would be covariant with respect to (frame-)transformations of an arbitrary gauge group.
The subtlety here is that the generalized coefficients of anholonomy $\Omega_{AB}{}^{C}$ defined in (\ref{newanhol}) rather
than the conventional ones appear in the last term of (\ref{naiveR}). Actually,  eq.~(\ref{gencomm})
does not determine whether (\ref{naivecur}) should contain the generalized coefficients of anholonomy
or the conventional ones, for in the commutator
(\ref{anhol}) the difference between the two is immaterial by virtue of the strong constraint (\ref{flatconstr}),
as we saw above. The choice made here is covariant under $\xi^{M}$ gauge
transformations, at the cost of violating the $GL(D)\times GL(D)$ covariance.

Next, we compute the failure of covariance in order to repair it in a second step, following
\cite{Siegel:1993th}.  The non-covariance can be inferred from the variation of the bare gauge field
in (\ref{finalcomm}),
 \be
 \begin{split}
  -\delta_{\Lambda} \omega^{D}{}_{[AB]}\,\nabla_{D}V_{C} \ &= \
  e^{D}\Lambda_{[A}{}^{F}\,{\cal G}_{B]F}\left(e_{D}V_{C}+\omega_{DC}{}^{E}V_{E}\right)
  \,+\,\text{covariant terms} \\
  \ &= \
  e^{E}\Lambda_{[A}{}^{F}\,{\cal G}_{B]F}\,\omega_{EC}{}^{D}V_{D}
   \,+\,\text{covariant terms}\;,
 \end{split}
\ee
where in the second line we relabeled indices and used the constraint (\ref{flatconstr}).
Since the non-covariance must be compensated by a non-covariant variation of $R$ we conclude
 \be\label{noncov}
  \delta_{\Lambda}R_{ABC}{}^{D} \ = \ -\left(e^{E}\Lambda_{[A}{}^{F}\right){\cal G}_{B]F} \,
  \omega_{EC}{}^{D}\,+\,\text{covariant terms}\;,
 \ee
which can also be checked directly with (\ref{naiveR}). 
We define now  a modified curvature tensor \cite{Siegel:1993th}
\be \label{modified_riemann}
 {\cal R}_{ABCD}^{\prime} \ \equiv \ \frac{1}{2} \Big[ R_{ABCD} + R_{CDAB} \Big]  + \frac{1}{4} \Big[ \omega_{ECD}\,
  \omega^{E}{}_{BA} +   \omega_{EAB} \,\omega^{E}{}_{DC} \Big] \, .
\ee
Using (\ref{noncov}) and (\ref{connvar}) it is straightforward to compute the non-covariant terms in its gauge variation,
 \begin{eqnarray}\nonumber
  \delta_{\Lambda}{\cal R}^{\prime}_{ABCD}
   &=& -\frac{1}{2}\left(e^{E}\Lambda_{A}{}^{F}\right){\cal G}_{BF}\,\omega_{E(CD)}
  -\frac{1}{2}\left(e^{E}\Lambda_{C}{}^{F}\right){\cal G}_{DF}\,\omega_{E(AB)}
   \\
   &=& \frac{1}{4}\left(e^{E}\Lambda_{A}{}^{F}\right){\cal G}_{BF}\,e_{E}{\cal G}_{CD}
   +\frac{1}{4}\left(e^{E}\Lambda_{C}{}^{F}\right){\cal G}_{DF}\,e_{E}{\cal G}_{AB} 
    \ \, =  \ \,  0\;,  \end{eqnarray}
where we used (\ref{metrsol}) and the constraint (\ref{flatconstr}). Thus, ${\cal R}^{\prime}$ transforms
covariantly under all symmetries. 
Since the proof of covariance requires the use of the metricity condition, ${\cal R}^{\prime}$
transforms only covariantly after imposing this constraint. This can, however, be relaxed by adding further
terms that are zero upon imposing the constraints. Specifically,  defining
 \be\label{offcovtensor}
  {\cal R}_{ABCD} \ = \ {\cal R}^{\prime}_{ABCD} -\frac{1}{4}\omega_{ECD}\,\nabla^{E}{\cal G}_{AB}
  -\frac{1}{4}\omega_{EAB}\,\nabla^{E}{\cal G}_{CD} \;,
 \ee
we obtain a tensor that is fully covariant independently of the constraints. In the remainder of this paper
we will assume that all constraints are satisfied, for which ${\cal R}={\cal R}^{\prime}$, unless stated differently.

In the rest of this section we examine the symmetry properties and identities of ${\cal R}_{ABCD}$.
We start with the original curvature $R_{ABC}{}^{D}$. It is manifestly antisymmetric in its first two
indices. It is not manifestly antisymmetric in its last two indices, but this follows nevertheless as
a consequence of the metricity condition. To see this we write (\ref{gencomm}) acting on a vector
with an upper index,
\be\label{commup}
 [\nabla_A , \nabla_B] V^D  \ = \ T _{AB}{}^E \nabla_E V^D - R_{ABC}{}^D V^C\;.
\ee
By the covariant constancy of the metric this is related to the commutator acting on a vector with a lower
index,
 \be
 \begin{split}
  {\cal G}^{CD} [\nabla_A , \nabla_B] V_C \ &=
  \ {\cal G}^{CD}  \Big( T _{AB}{}^E \nabla_E V_C + R_{ABC}{}^E V_E \Big) 
  \ = \ T _{AB}{}^E \nabla_E V^D + R_{AB}{}^D{}_{C}V^C \, .
 \end{split}
 \ee
Comparison with (\ref{commup}) then implies the antisymmetry in the last two indices.
Summarizing, $R$ has the following symmetries
 \be
  R_{ABCD} \ = \ -R_{BACD} \ = \ -R_{ABDC}\;.
 \ee
Moreover, since the gauge group is $GL(D)\times GL(D)$ the `off-diagonal' components in the group indices of
$R_{ABC}{}^{D}$, i.e., in the last two indices, are zero,
 \be
  R_{AB c\bar{d}} \ = \ R_{AB \bar{c}d} \ = \ 0\;,
 \ee
corresponding to the fact that the only non-zero connections are (\ref{nonzero}).

Next, we turn to the symmetry properties of ${\cal R}$. In general, the correction terms
proportional to the connections in (\ref{modified_riemann}) have no specific symmetry.
If we focus on off-diagonal $GL(D)\times GL(D)$ components, however, these extra terms vanish,
see (\ref{nonzero}), and so the antisymmetry properties of $R$ elevate to ${\cal R}$. For instance,
 \be
  {\cal R}_{a b \bar{c} d} \ = \ \frac{1}{2} \Big[ R_{a b \bar{c} d} + R_{\bar{c} d a b} \Big]
  \ = \ \frac{1}{2}R_{\bar{c} d a b}  \ = \ -\frac{1}{2}R_{\bar{c} d ba} \ = \
  -{\cal R}_{ba \bar{c} d} \, .
 \ee
The same conclusion applies to all other components that have precisely three unbarred or
three barred indices.

We close this section with a brief discussion of a curvature scalar that will be used in the next section
to define an action. The scalar that is obtained by tracing ${\cal R}$ turns out to be zero by virtue of the
constraints. Specifically, prior to imposing any constraints, one can prove (see appendix A1) 
that\footnote{We note that this expression differs from that in sec.~VIII of \cite{Siegel:1993th} because
of different conventions regarding symmetrization. Moreover, it differs by an overall factor and a relative 
factor in the last term.}
 \be\label{offscalar}
 {\cal R}_{AB}{}^{AB} \ = \ 2 \nabla_A \tilde{T}^A +  \tilde{T}_A{}^2 + \nabla_A \nabla_B  {\cal G}^{AB} 
 - \frac{1}{6} {\cal T}_{[ABC]}{}^2 - \frac{3}{8} \nabla_{(A} {\cal G}_{BC)}{}^2 
 \, .
 \ee
Each term vanishes separately after imposing the constraints, and therefore
 \be
   0 \ = \ {\cal R}_{AB}{}^{AB} \ = \
   {\cal R}_{ab}{}^{ab}  + {\cal R}_{\bar{a} \bar{b}}{}^{\bar{a} \bar{b}} \;.
 \ee
Thus, there is a unique way to define a (non-vanishing) scalar,
 \be\label{originalscalar}
   {\cal R} \ := \ -\frac{1}{2}{\cal R}_{ab}{}^{ab} \ = \ \frac{1}{2}{\cal R}_{\bar{a}\bar{b}}{}^{\bar{a}\bar{b}}\;,
 \ee
which by construction is a scalar under $\xi^{M}$ transformations and $GL(D)\times GL(D)$.

An expression for ${\cal R}$ that makes the invariance under $O(D,D)$ and frame transformations 
manifest is the following, 
 \begin{eqnarray}\label{explR} 
{\cal R} &=& -\left(\nabla^a \nabla_a d - \nabla^{\bar{a}} \nabla_{\bar{a}} d\right) 
- \frac{1}{2} \left(\nabla^a (e_{a}{}^M \nabla^{\bar{b}} e_{\bar{b} M} ) 
- \nabla^{\bar{a}} (e_{\bar{a}}{}^M \nabla^b e_{b M})  \right)   \\ \nonumber 
&& - \frac{1}{4} \left(e_{a}{}^M \nabla^b e^{\bar{c}}{}_M \,e^{aN}\nabla_{b} e_{\bar{c} N} -   e_{\bar{a}}{}^M \nabla^{\bar{b}} e^{c}{}_M \, e^{\bar{a}N}\nabla_{\bar{b}} e_{ c N}\right) \\ \nonumber
&& + \frac{1}{2 } \left(e_{\bar{c}}{}^M \nabla_a e_{b M}\,e^{\bar{c} N} \nabla^b e^a{}_N - e_{c}{}^M  \nabla_{\bar{a}} e_{\bar{b} M}\,e^{c N}  \nabla^{\bar{b}} e^{\bar{a}}{}_N \right)  \\ \nonumber
&& - \left(\nabla^a d \, (e_{a}{}^M \nabla^{\bar{b}} e_{\bar{b} M})  - \nabla^{\bar{a}} d \, (e_{\bar{a}}{}^M {\nabla}^{b} e_{b M}  )\right) - \left(\nabla^a d\, \nabla_a d - \nabla^{\bar{a}} d\, 
\nabla_{\bar{a}} d\right) \, ,
\end{eqnarray}
which will be verified in appendix A3. 

It is not manifest either from the definition (\ref{originalscalar}) or the explicit form (\ref{explR}) 
that the scalar curvature depends only on the connection components that have been
determined by the constraints. A somewhat lengthy calculation shows, however, that
${\cal R}$ can be written as\footnote{Again, this expression differs from that given in sec.~VIII of \cite{Siegel:1993th} because
of different conventions regarding antisymmetrization, but it also corrects a typo in the fourth term.}
  \be\label{finalscalar}
   {\cal R} \ = \ e_a \tilde{\Omega}^a + \frac{1}{2} \tilde{\Omega}_a{}^2  + \frac{1}{2} e_a e_b {\cal G}^{ab}   -
   \frac{1}{4} \Omega_{ a b \bar{c}} {}^2  -  \frac{1}{12} \Omega_{[abc]}{}^2
   + \frac{1}{8} e^a {\cal G}^{bc}\, e_b
   {\cal G}_{ac}\;,
 \ee
as we show in appendix A2. This proves that ${\cal R}$ is a well-defined function of the physical
fields.

\section{General action principle}\setcounter{equation}{0}
In this section we briefly introduce an Einstein-Hilbert like action principle based on the invariant 
curvature scalar discussed above, and derive Bianchi identities from its gauge invariance.

\subsection{Gauge invariant action}
Having the scalar ${\cal R}$ at our disposal we can define the following action principle
 \be\label{actionfinal}
   S \ = \ \int dxd\tilde{x}\,e^{-2d}\,{\cal R}\;,
 \ee
which, by virtue of $e^{-2d}$ transforming as a density, is manifestly invariant under all symmetries.

There are a number of conclusions that can be derived from this invariance. First, the variation
with respect to $d$ has to be a $GL(D)\times GL(D)$ invariant scalar and therefore it must be
proportional to ${\cal R}$ defined in (\ref{originalscalar}) \cite{Siegel:1993th}, which conclusion agrees with the results of \cite{Hohm:2010jy,Hohm:2010pp},
as we will show below. Second, the general variation with respect to $e_{A}{}^{M}$ is non-trivial
only in its off-diagonal component, in the following sense. Introducing a variation with both indices
flat,
 \be\label{Delta}
  \Delta e_{AB} \ := \ e_{B}{}^{M}\delta e_{A M}\;,
 \ee
we infer that the $GL(D)\times GL(D)$ transformations (\ref{GLtrans}) read
 \be
  \Delta e_{AB} \ = \ e_{B}{}^{M}\Lambda_{A}{}^{C}e_{CM} \ = \ 
  \Lambda_{A}{}^{C}{\cal G}_{BC} \ = \ \Lambda_{AB}\; . 
 \ee  
By the constraint  (\ref{Goff})
this implies
 \be
    \Delta e _{ab} \ = \ \Lambda_{ab}\;, \qquad
    \Delta e_{\bar{a}\bar{b}} \ = \ \Lambda_{\bar{a}\bar{b}}\;, \qquad
    \Delta e_{a\bar{b}} \ = \ -\Delta e_{\bar{b}a} \ = \ 0\;.
 \ee
Consequently, the local $GL(D)\times GL(D)$ symmetry of the action implies the `Bianchi identity' that
the diagonal parts of the field equations obtained by variation with respect to $\Delta e _{ab}$
and $\Delta e_{\bar{a}\bar{b}}$ vanish identically. Thus, the only non-trivial part of the field
equation is obtained by variation with respect to, say, $\Delta e_{a\bar{b}}$.
In total, the variation of (\ref{actionfinal}) can be written as
 \be\label{GenVar}
   \delta S \ = \ \int dxd\tilde{x}\,e^{-2d}\left(-2\delta d\,{\cal R}+
   \Delta e_{a\bar{b}}{\cal R}^{a\bar{b}}\right)\;,
 \ee
giving rise to the field equations
 \be
   {\cal R} \ = \ 0\;, \qquad {\cal R}_{a\bar{b}} \ = \ 0\;.
 \ee

Next we discuss some general properties of these tensors. As indicated by the suggestive notation
it is natural to assume that the `Ricci tensor'  ${\cal R}_{a\bar{b}}$ derived from (\ref{actionfinal}) 
indeed follows from contracting 
the covariant curvature tensor introduced above. There are two candidates, 
${\cal R}_{\bar{c}a\bar{b}}{}^{\bar{c}}$ and ${\cal R}_{c\bar{b}a}{}^{c}$. The explicit expression 
for the first is
 \begin{eqnarray}\label{Ric1}
  {\cal R}_{a\bar{b}} \ = \ 2{\cal R}_{\bar{c}a\bar{b}}{}^{\bar{c}} \ = \ R_{\bar{c}a\bar{b}}{}^{\bar{c}}
  &=&
  e_{\bar{c}}\omega_{a\bar{b}}{}^{\bar{c}}-e_{a}\omega_{\bar{c}\bar{b}}{}^{\bar{c}}
  +\omega_{\bar{c}\bar{b}}{}^{\bar{d}}\omega_{a\bar{d}}{}^{\bar{c}}
  -\omega_{a\bar{b}}{}^{\bar{d}}\omega_{\bar{c}\bar{d}}{}^{\bar{c}}
  -\Omega_{\bar{c}a}{}^{E}\omega_{E\bar{b}}{}^{\bar{c}} \\ \nonumber
  &=&
  e_{\bar{c}}\omega_{a\bar{b}}{}^{\bar{c}}-e_{a}
  \omega_{\bar{c}\bar{b}}{}^{\bar{c}}+\omega_{d\bar{b}}{}^{\bar{c}} \omega_{\bar{c}a}{}^{d}
  -\omega_{a\bar{b}}{}^{\bar{d}} \omega_{\bar{c}\bar{d}}{}^{\bar{c}}  \;,
 \end{eqnarray}
where the torsion constraint (\ref{TORsion}) has been used in the first line.
The second expression is given by
 \begin{eqnarray} \label{second_ricci}
{\cal R}_{\bar{b}a} \ = \ R_{ c \bar{b} a}{}^{ c} &=& e_{c}\omega_{\bar{b} a}{}^{c}-e_{\bar{b}}\omega_{c a}{}^{c}
  +\omega_{c a}{}^{d}\omega_{\bar{b} d}{}^{c}
  -\omega_{\bar{b} a }{}^{d}\omega_{c d}{}^{c}
  -\Omega_{c \bar{b} }{}^{E}\omega_{E a } {}^{c} \\ \nonumber
  &=&
  e_{c}\omega_{\bar{b} a}{}^{c}-e_{\bar{b}}\omega_{c a}{}^{c}
  +\omega_{\bar{d} a}{}^{c}\omega_{ c \bar{b} }{}^{\bar{d}}
  -\omega_{\bar{b} a }{}^{d}\omega_{c d}{}^{c}\;,
\end{eqnarray}
and we will confirm below that this is equivalent to (\ref{Ric1}). 
Writing out all connection components explicitly, the Ricci tensor can thus be written as
 \be
 {\cal R}_{a \bar{b}} \ = \ {\cal R}_{\bar{b}a} \ = \ 
 e_{\bar{b}} \tilde{\Omega}_a - e_c \Omega_{\bar{b} a}{}^{c} +  \Omega_{c \bar{b}}
 {}^{\bar{d}} \,\Omega_{\bar{d} a}{}^{c} - \Omega_{\bar{b} a}{}^c \,\tilde{\Omega}_{c}\;.
\ee
In sec.~4 we will prove that the curvature scalar, upon gauge fixing, reduces to the one of double 
field theory given in \cite{Hohm:2010jy}, and that the corresponding field equations for ${\cal E}_{ij}$ as 
determined in  \cite{Kwak:2010ew} give rise to the tensors in (\ref{Ric1}) or (\ref{second_ricci}),   thus
showing their equivalence. This proves that the tensors defined by the general variation (\ref{GenVar}) 
are indeed the curvature scalar and Ricci tensor.

\subsection{Covariant gauge variation and Bianchi identity}
In this subsection we derive a Bianchi identity from the invariance of (\ref{actionfinal}) under $\xi^{M}$ gauge
transformations. To this end it is convenient to first rewrite the gauge transformations in terms of the
$GL(D)\times GL(D)$ covariant derivatives.
For this we use the following form of the gauge transformation in terms of the C-bracket 
(c.f.~eqs. (3.29) and (3.30) in \cite{Hohm:2010pp})
 \be\label{Cbracketgauge}
  \delta_{\xi} e_{A}{}^{M} \ = \ \big[\xi,e_{A}\big]_{\rm C}^{M}+\frac{1}{2}
  \partial^{M}\big(e_{A}{}^{N}\xi_{N}\big)\;,
 \ee
and the fact that in the C bracket we can replace curved by flat indices if we use
the $GL(D)\times GL(D)$ covariant derivatives, i.e.,
 \begin{eqnarray}\label{Cbracketflat}
  \big[\xi,e_{A}\big]_{\rm C}^{B} &=& \xi^{C}\hat{\nabla}_{C}e_{A}{}^{B}-e_{A}{}^{C}\nabla_{C}\xi^{B}
  -\frac{1}{2}\xi_{C}\hat{\nabla}^{B}e_{A}{}^{C}+\frac{1}{2}e_{AC}\nabla^{B}\xi^{C} \\ \nonumber
  &=& -\xi^{C}\omega_{CA}{}^{B}-\nabla_{A}\xi^{B}
  +\frac{1}{2}\xi^{C}\omega^{B}{}_{AC}
  +\frac{1}{2}{\cal G}_{AC}\nabla^{B}\xi^{C}\;.
 \end{eqnarray}
Here we have to stress that the covariant derivatives in the first line do not act on the index $A$,
which we indicated by the notation $\hat{\nabla}$,
because $A$ is in (\ref{Cbracketgauge}) and (\ref{Cbracketflat}) only a `spectator' index.
Consequently, using $e_{A}{}^{B}\equiv e_{A}{}^{M}e_{M}{}^{B} =\delta_{A}{}^{B}$
and $e_{AC}\equiv e_{AM}e_{C}{}^{M}={\cal G}_{AC}$, we have
$\hat{\nabla}_{C}e_{A}{}^{B}=-\omega_{CA}{}^{B}$, from which the second equality follows.
Using (\ref{Cbracketflat}) in (\ref{Cbracketgauge}) we obtain
 \begin{eqnarray}
  \delta_{\xi}e_{A}{}^{M} &=& e_{B}{}^{M}\big[\xi,e_{A}\big]_{\rm C}^{B}
  +\frac{1}{2}\partial^{M}\xi_{A} \\ \nonumber
  &=& -\xi^{C}\omega_{CA}{}^{B}\,e_{B}{}^{M}
  -e_{B}{}^{M}\nabla_{A}\xi^{B}
  +\frac{1}{2}e_{B}{}^{M}\omega^{B}{}_{AC}\xi^{C}
  +\frac{1}{2}e_{B}{}^{M}\nabla^{B}\xi_{A}
  +\frac{1}{2}\partial^{M}\xi_{A}\;.
 \end{eqnarray}
The third and last term combine into a covariant derivative, which in turn combines with the fourth term. 
Moreover, the first term can be viewed as a field-dependent $GL(D)\times GL(D)$ transformation
with parameter $\Lambda_{A}{}^{B}=-\xi^{C}\omega_{CA}{}^{B}$ and can thus be discarded.
Therefore,  the final form reads
 \be\label{deltacov}
   \delta_{\xi}e_{A}{}^{M} \ = \ -e_{B}{}^{M}\left(\nabla_{A}\xi^{B}-\nabla^{B}\xi_{A}\right)
      \;,
 \ee
or, in terms of the variation (\ref{Delta}),
 \be\label{Deltacov}
   \Delta e_{AB} \ = \  \nabla_{B}\xi_{A}-\nabla_{A}\xi_{B}\;.
 \ee

For the dilaton one finds from (\ref{finalgtINTRO})
 \be
 \begin{split}\label{covd}
   \delta_{\xi}d \ &= \ \xi^{M}\partial_{M}d-\frac{1}{2}\partial_{M}\xi^{M} \ = \
   \xi^{A}e_{A}d-\frac{1}{2}\partial_{M}\left(\xi^{A}e_{A}{}^{M}\right) \\
   \ &= \ -\frac{1}{2}e_{A}\xi^{A}+\frac{1}{2}\xi^{A}\left(-\partial_{M}e_{A}{}^{M}+2e_{A}d\right)
   \ = \  -\frac{1}{2}\left(e_{A}\xi^{A}-\omega_{BA}{}^{B}\xi^{A}\right)\\
   \ &= \ -\frac{1}{2}\nabla_{A}\xi^{A}\;,
  \end{split}
  \ee
where we used (\ref{tracepart}) in the second line.

We can now read off the Bianchi identity following from the gauge invariance of (\ref{actionfinal}).
Using (\ref{Deltacov}) and (\ref{covd}) in (\ref{GenVar}) we infer
 \be
 \begin{split}
  0 \ &= \ \delta_{\xi}S \ = \ \int dx d\tilde{x}\,e^{-2d}\left(\left(\nabla_{a}\xi^{a}+\nabla_{\bar{a}}\xi^{\bar{a}}\right)
  {\cal R}+\left(\nabla_{\bar{b}}\xi_{a}-\nabla_{a}\xi_{\bar{b}}\right){\cal R}^{a\bar{b}}\right) \\
  \ &= \ -\int dxd\tilde{x}\,e^{-2d}\left(\xi^{a}\left(\nabla_{a}{\cal R}+\nabla^{\bar{b}}{\cal R}_{a\bar{b}}\right)
  +\xi^{\bar{a}}\left(\nabla_{\bar{a}}{\cal R}-\nabla^{b}{\cal R}_{b\bar{a}}\right)\right)\;,
 \end{split}
 \ee
which implies the Bianchi identities \cite{Siegel:1993th}
 \be\label{Bianchi4}
     \nabla_{a}{\cal R}+\nabla^{\bar{b}}{\cal R}_{a\bar{b}} \ = \ 0\;, \qquad
     \nabla_{\bar{a}}{\cal R}-\nabla^{b}{\cal R}_{b\bar{a}} \ = \ 0\;.
 \ee
These are equivalent to similar Bianchi identities derived from the double field theory,
as we will show in the next section, and reduce to the usual Bianchi identities for $R_{ij}$ and 
$H_{ijk}$ when $\tilde{\partial}=0$  \cite{Kwak:2010ew}.

\section{Relation to formulation with ${\cal E}_{ij}$}\setcounter{equation}{0}
Here we start the detailed `re-derivation' of the double field theory formulations reviewed in the introduction
from Siegel's geometrical formalism. We identify the `non-symmetric'
metric ${\cal E}_{ij}$ as components of $e_{A}{}^{M}$ after a particular gauge fixing. This allows us to 
study the non-linear realization of the $O(D,D)$ symmetry and to 
find a rather direct relation between the action (\ref{THEActionINTRO}) and the geometrical 
Einstein-Hilbert like action (\ref{actionfinal}).

\subsection{Gauge choice}
One way to identify ${\cal E}_{ij}$ in the frame-like formalism is to gauge-fix the
local $GL(D)\times GL(D)$ symmetry by setting the components $e_{a}{}^{i}$ and $e_{\bar{a}}{}^{i}$ in
(\ref{Egauge}) equal to the unit matrix (assuming certain invertibility properties). Taking the constraint
(\ref{Goff}) into account, the remaining components are then parametrized by a general $D\times D$ matrix
which we identify with ${\cal E}_{ij}$,\footnote{An alternative definition of
${\cal E}_{ij}$ in terms of the frame fields which is
$GL(D)\times GL(D)$ covariant and does not require a gauge fixing has been given in \cite{Hohm:2010pp}. For our present purposes, however,
we find it more convenient to use the gauge fixed form (\ref{Egauge2}).}
 \be\label{Egauge2}
   e_{A}{}^{M} \ = \ \begin{pmatrix} e_{ai} &  e_{a}{}^{i} \\ e_{\bar{a}i} & e_{\bar{a}}{}^{i} \end{pmatrix}
   \ = \ \begin{pmatrix} -{\cal E}_{ai} &   \delta_{a}{}^{i} \\ {\cal E}_{i\bar{a}} & \delta_{\bar{a}}{}^{i} \end{pmatrix}\;.
 \ee
In this gauge, the `space-time' indices $i,j,\ldots$ can be identified with the frame indices of either
$GL(D)$ factor via the trivial vielbeins $\delta_{a}{}^{i}$ or $\delta_{\bar{a}}{}^{i}$. The calligraphic derivatives (\ref{groihffkdf}) then coincide with the `flattened' partial derivatives (\ref{flatder}),
 \be\label{callderflat}
  e_{a} \ = \ e_{a}{}^{M}\partial_{M} \ = \ \partial_{a}-{\cal E}_{ai}\tilde{\partial}^{i} \ \equiv \ {\cal D}_{a}\;, \qquad
  e_{\bar{a}} \ = \ e_{\bar{a}}{}^{M}\partial_{M} \ = \ \partial_{\bar{a}}+{\cal E}_{i\bar{a}}\tilde{\partial}^{i} \ \equiv \
  \bar{\cal D}_{\bar{a}}\;.
 \ee
Moreover, the metric $g_{ij}={\cal E}_{(ij)}$ can be identified with either of the two `tangent space' metrics
 \be\label{gs}
  g_{ab} \ = \ -\frac{1}{2}e_{a}{}^{M}e_{b}{}^{N}\eta_{MN}\;, \qquad
  g_{\bar{a}\bar{b}} \ = \ \frac{1}{2}e_{\bar{a}}{}^{M}e_{\bar{b}}{}^{N}\eta_{MN}\;,
 \ee
as one may verify directly from (\ref{Egauge2}). From this it follows that (\ref{flatmetric}) is given by
 \be
  {\cal G}_{AB} \ = \  \begin{pmatrix} -2g_{ab} &  0 \\ 0 & 2g_{\bar{a}\bar{b}} \end{pmatrix}\;.
 \ee
The relative factors of $\pm 2$ appearing here lead, after the gauge fixing (\ref{Egauge2}) and the 
corresponding identification of indices, to an ambiguity regarding the contraction of indices.
We will follow the convention that contractions are done with respect 
to the tangent space metric ${\cal G}_{AB}$ when the indices are letters from the beginning 
of the latin alphabet (i.e.,~either $a,b\ldots$ or $\bar{a},\bar{b},\ldots$), and that contractions 
are only done with respect to $g_{ij}$ if the indices are letters from the middle of the latin 
alphabet ($i,j,\ldots$). 

For the comparison with the action (\ref{THEActionINTRO}) it is instructive to re-interpret derivatives
like ${\cal D}_{i}{\cal E}_{jk}$ in a more covariant way. Specifically, in analogy to the modified variation (\ref{Delta}), we can write this as
 \be
  {\cal D}_{a}{\cal E}_{b\bar{c}} \ = \ e_{b}{}^{M}e_{a}e_{\bar{c}M} \ = \ -e_{\bar{c}}{}^{M}e_{a}e_{bM}\;.
 \ee
This follows from the gauge fixed forms (\ref{Egauge2}) and (\ref{callderflat}), and is
manifestly $O(D,D)$ invariant. Remarkably, it can also be made manifestly $GL(D)\times GL(D)$
invariant by observing that in
 \be\label{connectionrel}
   e_{b}{}^{M}\nabla_{a}e_{\bar{c}M} \ = \ e_{b}{}^{M}\big(e_{a}e_{\bar{c}M}
   +\omega_{a\bar{c}}{}^{\bar{d}} e_{\bar{d}M}\big)
 \ee
the connection term is zero by the constraint (\ref{Goff}).  The same conclusion applies to the barred
derivative $e_{\bar{a}}=\bar{\cal D}_{\bar{a}}$, and so we find in total the following identifications
 \be\label{calEcov}
 \begin{split}
  {\cal D}_{a}{\cal E}_{b\bar{c}} \ &\equiv \ e_{b}{}^{M}\nabla_{a}e_{\bar{c}M}
  \ = \ -e_{\bar{c}}{}^{M}\nabla_{a}e_{bM}\;, \\[0.5ex]
  \bar{\cal D}_{\bar{a}}{\cal E}_{b\bar{c}} \ &\equiv \ e_{b}{}^{M}\nabla_{\bar{a}}e_{\bar{c}M}
   \ = \ -e_{\bar{c}}{}^{M}\nabla_{\bar{a}}e_{bM}\;,
 \end{split}
 \ee
which are manifestly covariant with respect to $O(D,D)$ and tangent space transformations.

In the following we will examine how the $O(D,D)$ duality symmetry is realized after
this gauge fixing. Acting with a general $O(D,D)$ transformation on (\ref{Egauge2}) violates
the gauge condition and thus requires a compensating $GL(D)\times GL(D)$ transformation.
In order to determine the transformation that restores the form of the vielbein (\ref{Egauge2}),
we consider a finite $O(D,D)$ and $GL(D)$ transformation,
 \be
   e_{a}{}^{M\prime}(X^{\prime}) \ = \ h^{M}{}_{N}\,(M^{-1}(X))_{a}{}^{b}\,e_{b}{}^{N}(X)\;.
 \ee
Here we denoted the $GL(D)$ matrix by $M^{-1}$ for later convenience, and $h$ is the $O(D,D)$
matrix in (\ref{ODDmatrix}), i.e., whose components read
 \be
   h^{M}{}_{N} \ = \ \begin{pmatrix} h_{i}{}^{j} &  h_{ij} \\
   h^{ij} & h^{i}{}_{j} \end{pmatrix} \ = \
   \begin{pmatrix} a_{i}{}^{j} &  b_{ij} \\ c^{ij}  & d^{i}{}_{j} \end{pmatrix}\;.
  \ee
Applied to the gauge fixed component we find
 \be\label{gaugecon1}
  e_{a}{}^{i\prime} \ = \ (M^{-1})_{a}{}^{b}\left(h^{i}{}_{j}e_{b}{}^{j}+h^{ij}e_{bj}\right)
  \ = \ \big(M^{-1}(d^{t}-{\cal E} c^{t})\big)_{a}{}^{i} \ = \ \delta_{a}{}^{i} \;,
\ee
where we used matrix notation and suppressed the $X$-dependence.
The last equation expresses the condition that the
gauge fixing condition be preserved.  Analogously, one finds for the other component
 \be\label{gaugecon2}
   e_{\bar{a}}{}^{i\prime} \ = \ (\bar{M}^{-1})_{\bar{a}}{}^{\bar{b}} \left(h^{i}{}_{j} e_{\bar{b}}{}^{j}+
   h^{ij}e_{\bar{b}j}\right) \ = \ \left(\bar{M}^{-1}(d^t+{\cal E}^{t}c^{t})\right)_{\bar{a}}{}^{i}
   \ = \ \delta_{\bar{a}}{}^{i}\;,
 \ee
where we denoted the matrix corresponding to the second $GL(D)$ factor by $\bar{M}^{-1}$.
The two conditions  (\ref{gaugecon1}) and (\ref{gaugecon2}) thus determine the compensating
$GL(D)\times GL(D)$ transformations uniquely in terms of $c$ and $d$,
 \be\label{MbarM}
   M(X) \ = \ d^t-{\cal E}(X)c^t \;, \qquad \bar{M}(X) \ = \ d^t+{\cal E}^t(X)c^t\;,
 \ee
which are both $X$-dependent through their dependence on ${\cal E}_{ij}$.
Finally, using this form of the compensating gauge transformations it is straightforward to verify that
${\cal E}_{ij}$ transforms under $O(D,D)$ in the required non-linear representation according to  (\ref{ODDaction}).

With the above analysis of  the non-linear realization of $O(D,D)$ we have in fact recovered the
formalism that has been used in \cite{Hohm:2010jy} (extending the background-dependent
formalism in \cite{Kugo:1992md,Hull:2009mi})
in order to prove the $O(D,D)$ invariance of (\ref{THEActionINTRO}).  More precisely, in this
formalism every index is thought of either as an unbarred or barred index and
to transform, accordingly, either under $M$ or $\bar{M}$ in (\ref{MbarM}). For instance, we have just verified
that the calligraphic derivatives (\ref{callderflat}) transform with $M$ or $\bar{M}$, respectively.
Moreover, due to the manifestly $O(D,D)$ and $GL(D)\times GL(D)$ covariant rewriting of the calligraphic derivatives of ${\cal E}$ in (\ref{calEcov}), it follows that after gauge fixing
 \be
   {\cal D}_{a}{\cal E}_{b\bar{c}} \ = \ M_{a}{}^{d}\,M_{b}{}^{e}\,\bar{M}_{\bar{c}}{}^{\bar{f}}\,
   {\cal D}_{d}^{\prime}{\cal E}^{\prime}_{e\bar{f}}\;, \qquad
   \bar{\cal D}_{\bar{a}}{\cal E}_{b\bar{c}} \ = \ \bar{M}_{\bar{a}}{}^{\bar{d}}\,M_{b}{}^{e}\,
   \bar{M}_{\bar{c}}{}^{\bar{f}}\,\bar{\cal D}^{\prime}_{\bar{d}}{\cal E}^{\prime}_{e\bar{f}}\;.
 \ee
Thus, we can think of the first index on ${\cal E}$ (under ${\cal D}$ or $\bar{\cal D}$) as unbarred and the
second index as barred. 
From the definition (\ref{gs}) we infer that the indices on $g$ can be thought of either as both barred or
both unbarred, because $g$ can be viewed as a tensor either of the left 
$GL(D)$ or the right $GL(D)$ such that it  transforms after gauge fixing as 
 \be
  g_{\bar{a}\bar{b}} \ = \ \bar{M}_{\bar{a}}{}^{\bar{c}}\,\bar{M}_{\bar{b}}{}^{\bar{d}}\,g^{\prime}_{\bar{c}\bar{d}}
  \;, \qquad
  g_{ab} \ = \ M_{a}{}^{c}\, M_{b}{}^{d}\,g^{\prime}_{cd}\;, 
 \ee
and similarly for the inverse.  
The $O(D,D)$ invariance of the action is then a consequence of
the fact, which one may easily confirm by inspection of (\ref{THEActionINTRO}), that only like-wise indices are contracted  \cite{Hohm:2010jy}.

\subsection{$O(D,D)$ covariant derivatives and gauge variation}\label{ODDder}
In the previous subsection we have seen that in the formulation using ${\cal E}_{ij}$ the $O(D,D)$
transformations are governed by the matrices $M$ and $\bar{M}$ in (\ref{MbarM}).  Since these
matrices are $X$-dependent, it follows that derivatives of objects that transform `covariantly' with
$M$ and $\bar{M}$ according to their index structure are in general not covariant in the same sense.
This led ref.~\cite{Hohm:2010jy} to introduce `$O(D,D)$ covariant derivatives' --- despite $O(D,D)$ 
being a \textit{global} symmetry with constant parameters.
There are two types of covariant derivatives,
$\nabla_{i}(\Gamma)$ and $\bar{\nabla}_{i}(\Gamma)$, i.e., unbarred and barred, and various connections
$\Gamma$ depending on the index structure of the object on which the derivative acts.
Here we indicate the dependence on the connections explicitly, in order to distinguish these
`covariant' derivatives from the $GL(D)\times GL(D)$ covariant derivatives introduced before.

Since we have here realized the global non-linear $O(D,D)$ transformations according to
$M$ and $\bar{M}$ through compensating $GL(D)\times GL(D)$ transformations, it is natural to assume
that, after gauge fixing, the $GL(D)\times GL(D)$ covariant derivatives are related to the
`$O(D,D)$ covariant derivatives' of  \cite{Hohm:2010jy}.
This indeed turns out to be the case, and so we are able to give a more conventional
interpretation of these covariant derivatives.

As a first test of this relation we reproduce a manifestly $O(D,D)$ covariant form of the $\xi^{M}$ gauge
transformations that  has been found in \cite{Hohm:2010jy}. Specifically,
introducing the following change of basis for the gauge parameters (which is suggested by the
gauge structure in string field theory \cite{Hull:2009mi}),
 \be\label{SFTbasis}
   \eta_{i} \ = \ -\tilde{\xi}_{i}+{\cal E}_{ij}\xi^{j}\;, \qquad
   \bar{\eta}_{i} \ = \ \tilde{\xi}_{i}+\xi^{j}{\cal E}_{ji}\;,
 \ee
the gauge transformations  (\ref{finalgtINTRO}) take the remarkable form
 \be\label{delcalE}
  \delta{\cal E}_{ij} \ = \ \nabla_{i}(\Gamma)\bar{\eta}_{j}+\bar{\nabla}_{j}(\Gamma)\eta_{i}\;.
\ee
The corresponding result using the $GL(D)\times GL(D)$
connections follows almost immediately. First, the flattened gauge parameters
 \be\label{etabase}
  \eta_{a} \ := \ -\xi_{a} \ \equiv \ -e_{a}{}^{M}\xi_{M}\;, \qquad
  \bar{\eta}_{\bar{a}} \ := \ \xi_{\bar{a}} \ \equiv \ e_{\bar{a}}{}^{M}\xi_{M}\;,
 \ee
coincide with (\ref{SFTbasis}) upon using (\ref{Egauge2}). Moreover, after the gauge fixing (\ref{Egauge2}),
any variation of ${\cal E}$ coincides with the $\Delta$ variation in (\ref{Delta}),
\be\label{DeltaE}
  \delta {\cal E}_{a\bar{b}} \ = \ \Delta e_{\bar{b}a} \ = \ e_{a}{}^{M}\,\delta e_{\bar{b}M}
    \ = \ e_{a}{}^{i}\, \delta e_{\bar{b}i} + e_{ai}\, \delta e_{\bar{b}}{}^{i}\;.
 \ee
This follows because the last term is zero by the gauge fixing condition. More precisely, for 
the $\xi^{M}$ gauge variation this term will vanish by a compensating frame rotation that restores
the chosen gauge. The advantage of using the $\Delta$ variation is that this compensating
transformation need not be determined explicitly.
Applying now (\ref{Deltacov}) one finds in the basis (\ref{etabase})
 \be\label{finalgauge}
  \delta  {\cal E}_{a\bar{b}} 
  \ = \  \nabla_{a}\bar{\eta}_{\bar{b}}+\nabla_{\bar{b}}\eta_{a}\;,
 \ee
which agrees with (\ref{delcalE}), using that after gauge fixing the indices $i,j,\ldots$ can be identified
with the flat indices.

We note in passing that the original form (\ref{finalgtINTRO}) of the gauge transformations also 
follows easily by use of the $\Delta$ variation as in (\ref{DeltaE}), 
 \be
 \begin{split}
  \delta {\cal E}_{a\bar{b}} \ &= \ \Delta e_{\bar{b}a} \ = \ e_{a}{}^{M}\delta e_{\bar{b}M} 
  \ = \ 
  e_{a}{}^{M}\left(\xi^{N}\partial_{N}e_{\bar{b}M}+(\partial_{M}\xi^{N}-\partial^{N}\xi_{M})e_{\bar{b}N}\right)
  \\[0.3ex]
  \ &= \ \xi^{N}\partial_{N}{\cal E}_{a\bar{b}}+{\cal D}_{a}\xi^{N} e_{\bar{b}N}-\bar{\cal D}_{\bar{b}}\xi_{M}
  e_{a}{}^{M} \\[0.3ex]
  \ &= \ 
  \xi^{N}\partial_{N}{\cal E}_{a\bar{b}}+{\cal D}_{a}\tilde{\xi}_{j}\, e_{\bar{b}}{}^{j}+{\cal D}_{a}\xi^{j} e_{\bar{b}j}
  -\bar{\cal D}_{\bar{b}}\tilde{\xi}_{j}\,e_{a}{}^{j}-\bar{\cal D}_{\bar{b}}\xi^{j} e_{aj}\\[0.3ex]
  \ &= \ \xi^{N}\partial_{N}{\cal E}_{a\bar{b}}+{\cal D}_{a}\tilde{\xi}_{\bar{b}}
  +{\cal D}_{a}\xi^{j} {\cal E}_{j\bar{b}}
    -\bar{\cal D}_{\bar{b}}\tilde{\xi}_{a}+\bar{\cal D}_{\bar{b}}\xi^{j}
   {\cal E}_{aj}\;.
 \end{split}
 \ee
Here we used (\ref{callderflat}) in the second line and the gauge fixed form (\ref{Egauge2}) 
in the last line, where we again identified indices.  Thus we have derived (\ref{finalgtINTRO}) 
from the fundamental gauge transformation of the vielbein, as in \cite{Hohm:2010pp}, 
but without invoking the compensating frame rotation explicitly.

The previous results show that the `$O(D,D)$ covariant derivatives' coincide with the
$GL(D)\times GL(D)$ covariant derivatives after gauge fixing, at least when acting on $\eta$ and
$\bar{\eta}$ as in (\ref{finalgauge}). The complete set of connections $\Gamma$ is not fixed by $O(D,D)$ covariance
and therefore have been given in \cite{Hohm:2010jy} only provisionally. Here we display for
completeness their relation after gauge fixing, 
  \begin{eqnarray}\label{Gammacor}
    \omega_{i \bar{j}}{}^{\bar{k}} &=& -\frac{1}{2} g^{kl} \Big( {\cal D}_i {\cal E}_{lj}
    + \bar{\cal D}_j {\cal E}_{il} - \bar{\cal D}_l {\cal E}_{ij}  \Big) \ = \ - \Gamma_{i \bar{j}}^{\bar{k}} \, ,
     \nonumber \\
     \omega_{\bar{i} j}{}^{k} &=& -\frac{1}{2} g^{kl} \Big( \bar{\cal D}_i {\cal E}_{jl}
     + {\cal D}_j {\cal E}_{l i} - {\cal D}_l {\cal E}_{ji}  \Big) \ = \  - \Gamma_{\bar{i} j}^{k} \, ,
     \nonumber \\
     \omega_{ji}{}^{j} &=& \frac{1}{2} (\bar{\cal D}^{j} - {\cal D}^{j}) {\cal E}_{ij}
     + 2 {\cal D}_i d \ = \ - \Gamma_{ji}^{j} + \frac{1}{2} \bar{\cal D}^{j}{\cal E}_{ij} + 2 {\cal D}_i d \, ,
     \nonumber \\
      \omega_{\bar{j} \bar{i}}{}^{\bar{j}} &=&  \frac{1}{2}  ({\cal D}^{j} - \bar{\cal D}^{j}) {\cal E}_{ji}
      + 2 \bar{\cal D}_i d  \ = \
      - \Gamma_{\bar{j} \bar{i}}^{\bar{j}} + \frac{1}{2}  {\cal D}^{j} {\cal E}_{ji} + 2 \bar{\cal D}_i d  \, .
 \end{eqnarray}
We see that they are equivalent in the `off-diagonal' parts but differ in the trace parts.
In fact, it has already been noted, c.f.~the discussion around eq.~(4.13) in
\cite{Hohm:2010jy}, that modifying the definition as suggested by (\ref{Gammacor}) 
would have the advantage of simplifying the gauge transformation of $d$ in that
 \be\label{deldcov}
  \delta d \ = \ -\frac{1}{4}\nabla_{i}\eta^{i}-\frac{1}{4}\bar{\nabla}_{i}\bar{\eta}^{i}\;.
 \ee
Here we see that this is a direct consequence of (\ref{covd}), where we recall that according to 
our index conventions $g$ rather than ${\cal G}$ is used to raise indices in (\ref{deldcov}), 
and that there is a 
relative sign in the definition (\ref{etabase}) of $\eta_{i}$.
In \cite{Hohm:2010jy}, however,
there was no justification from symmetry arguments for this modification, but here we see it
emerging naturally from Siegel's frame formalism.

Given the precise correspondence between the $O(D,D)$ and $GL(D)\times GL(D)$
connections, we have verified that the curvature scalar and Ricci tensor of Siegel's formalism
agree with the corresponding expressions obtained in \cite{Hohm:2010jy} and \cite{Kwak:2010ew}
(for the Ricci tensor see appendix A4).
More precisely, the scalar curvature constructed from Siegel's frame formalism is $\tfrac{1}{4}$ times 
${\cal R}({\cal E},d)$  as given in \cite{Hohm:2010jy}. Taking this factor as well as the relative 
factors of $\pm\tfrac{1}{2}$ in (\ref{gs}) into account, the Bianchi identities (\ref{Bianchi4}) reduce
to
 \be
  \nabla^{i}{\cal R}_{ij}+\frac{1}{2}\bar{\cal D}_{j}{\cal R}({\cal E},d) \ = \ 0\;, \qquad
  \bar{\nabla}^{j}{\cal R}_{ij}+\frac{1}{2}{\cal D}_{i}{\cal R}({\cal E},d) \ = \ 0\;,
 \ee
which agree with   \cite{Kwak:2010ew}.

Starting from the expression (\ref{explR}) for the scalar curvature we can actually immediately 
compare with the double field theory action   
(\ref{THEActionINTRO}) in terms of ${\cal E}_{ij}$. Using that the covariant derivatives allow for 
partial integration in presence of the dilaton density, we infer that the first line in (\ref{explR})
contributes only total derivatives under an integral, and thus the resulting Lagrangian
is equivalent to  
\be\label{doublecov}
 \begin{split}
  {\cal L}^{\prime} \ = \ e^{-2d}\Big(&-\frac{1}{2}e^{aM}\nabla^{b}e^{\bar{c}}{}_{M}\,
  e_{a}{}^{N}\nabla_{b}e_{\bar{c}N}
  +\frac{1}{2}e_{\bar{c}}{}^{M}\nabla_{a}e_{bM}\,e^{\bar{c}N}\nabla^{b}e^{a}{}_{N}
  -\frac{1}{2}e_{c}{}^{M}\bar{\nabla}_{\bar{a}}e_{\bar{b}M}\,e^{cN}\bar{\nabla}^{\bar{b}}e^{\bar{a}}{}_{N} 
     \\
  &-\nabla^{a}d\,e_{a}{}^{M}\bar{\nabla}^{\bar{b}}e_{\bar{b}M}
  +\bar{\nabla}^{\bar{a}}d\, e_{\bar{a}}{}^{M}
  \nabla^{b}e_{b}{}_{M}-2\nabla^{a}d\,\nabla_{a}d\Big)\;. 
 \end{split}
 \ee
Taking into account the relation    
(\ref{gs}) between $g$ and  the tangent space metric, and using that the latter is covariantly constant,
it then immediately follows by virtue of the 
identifications (\ref{calEcov}) that this agrees with (\ref{THEActionINTRO}) up to the 
overall factor of $4$.

\section{Relation to formulation with ${\cal H}^{MN}$}\setcounter{equation}{0}
In this section we introduce the formulation in terms of the generalized metric ${\cal H}^{MN}$ from 
the point of view of the frame formalism and express the scalar curvature and thus the action in terms 
of this variable. Finally, we briefly discuss Christoffel connections that are 
introduced via a vielbein postulate.

\subsection{Gauge choice and generalized coset formulation}
We next identify the generalized metric and the corresponding formulation (\ref{Hactionx}) in the
geometrical frame formalism. In general, one can define ${\cal H}^{MN}$ in terms of the frame field
through \cite{Hohm:2010pp}
 \be\label{Hroot}
  {\cal H}^{MN} \ = \ 2{\cal G}^{\bar{a}\bar{b}}\,e_{\bar{a}}{}^{M}e_{\bar{b}}{}^{N}-\eta^{MN} \ = \
  -2{\cal G}^{ab}\,e_{a}{}^{M}e_{b}{}^{N}+\eta^{MN}\;,
 \ee
where the second equation is a consequence of the definition (\ref{flatmetric}) and  the constraint
(\ref{Goff}).  
The generalized metric is a constrained field in that 
 \be\label{Hconst}
  {\cal H}^{MK}{\cal H}_{KN} \ = \ \delta^{M}{}_{N}\;,
 \ee
where the indices are lowered, as usual, with $\eta_{MN}$. In the standard  
parametrization (\ref{genmetric}) this can be checked by a direct computation. Here, it can be verified 
with either one of the definitions in (\ref{Hroot}). We note, however, that if we use for the first ${\cal H}$ in (\ref{Hconst}), say, the first expression in (\ref{Hroot}) and for the second ${\cal H}$ the second expression, then 
the constraint (\ref{Goff}) is required in order to verify this.  
 
For later use we note that (\ref{Hroot}) implies for the flattened components of the generalized 
metric 
 \be\label{Hflat}
  {\cal H}^{AB} \ = \ {\cal H}^{MN}\,e_{M}{}^{A}\,e_{N}{}^{B} \ = \ 
  \begin{pmatrix} -{\cal G}^{ab} &  0 \\ 0 & {\cal G}^{\bar{a}\bar{b}} \end{pmatrix}\;,
 \ee
where again (\ref{Goff}) has been used.   
 
In the following, we find
it convenient to fix the $GL(D)\times GL(D)$ symmetry by setting the tangent space metric (\ref{flatmetric})  to
 \be\label{Gfix}
   {\cal G}_{AB} \ = \  \begin{pmatrix} -\delta_{ab} &  0 \\ 0 & \delta_{\bar{a}\bar{b}} \end{pmatrix}\;.
 \ee
This implies $g_{ab}=\tfrac{1}{2}\delta_{ab}$ and
$g_{\bar{a}\bar{b}} =\tfrac{1}{2}\delta_{\bar{a}\bar{b}}$ from the definition (\ref{gs}) and also 
${\cal H}_{AB}=\delta_{AB}$ from (\ref{Hflat}). 
This leaves a residual local $O(D)\times O(D)$
symmetry. Therefore, the resulting formulation can be viewed as a generalized coset model
based on $O(D,D)/(O(D)\times O(D))$ \cite{Hohm:2010pp}.  In fact, from (\ref{Gfix}) we conclude with (\ref{flatmetric}) that $e_{A}{}^{M}$ is an $O(D,D)$ element
(up to a similarity
transformation) in that it transforms the $O(D,D)$ metric $\eta$ into the
$O(D,D)$ metric, but written in the form (\ref{Gfix}). Thus, $e$ can be viewed as a group-valued
coset representative with a local $O(D)\times O(D)$ action from the left.
Moreover, (\ref{Hflat}) implies 
 \be\label{Hcoset}
    {\cal H}^{MN} \ = \ \delta^{AB}\,e_{A}{}^{M}\,e_{B}{}^{N}\;,
 \ee
and so ${\cal H}$ can be viewed as the $O(D)\times O(D)$ invariant combination $e^{t}e$.
For completeness we record that the form of the coset representative that leads to the standard
parametrization (\ref{genmetric}) for ${\cal H}^{MN}$ according to (\ref{Hcoset}) is given by
 \be\label{coset}
   e_{A}{}^{M} \ = \ \frac{1}{\sqrt{2}} \begin{pmatrix} v_{ai}+b_{ij}v_{a}{}^{j} & v_{a}{}^{i} \\
   -v_{\bar{a}i}+b_{ij}v_{\bar{a}}{}^{j} & v_{\bar{a}}{}^{i} \end{pmatrix}\;,
 \ee
where $v_{i}{}^{a}$ is the conventional vielbein for the metric $g_{ij}$, i.e., $g_{ij} =  v_{i}{}^{a}v_{ja}$,
with inverse $v_{a}{}^{i}$. We recall that an explicit parametrization like this requires a further gauge fixing of the local $O(D)\times O(D)$ symmetry.

\subsection{Scalar curvature}
Next, we prove that the Ricci scalar (\ref{finalscalar}) reduces upon the gauge fixing (\ref{Gfix}) to the
function ${\cal R}({\cal H},d)$ given in \cite{Hohm:2010pp}, and thus that the actions in (\ref{actionfinal})
and (\ref{masteractionINTRO}) are equivalent.
This proof simplifies due to the fact that we have chosen a gauge in which ${\cal G}_{AB}$ is constant,
such that we can freely 
raise and lower indices $a,b,\ldots$ and $\bar{a},\bar{b},\ldots$ under derivatives.
Thus, it implies relations like
 \be\label{simplyf}
   f_{ABC} \ = \ \big(e_{A}e_{B}{}^{M}\big)e_{CM} \ = \  -e_{B}{}^{M}\big(e_{A}e_{CM}\big)
   \ = \ 
   -f_{ACB}\;,
 \ee
which we will use frequently below.
Moreover, the scalar curvature (\ref{finalscalar}) then reduces to
 \be\label{Rredux}
   {\cal R} \ = \ e_{a}\tilde{\Omega}^{a}+\frac{1}{2}\tilde{\Omega}_{a}^2-\frac{1}{4}\Omega_{ab\bar{c}}{}^{2}
   -\frac{1}{12}\Omega_{[abc]}{}^{2}\;.
 \ee

We first evaluate the dilaton-dependent terms, which originate only from the first two terms. Using 
(\ref{tracepart}) we find
 \begin{eqnarray}
   e_{a}\tilde{\Omega}^{a}+\frac{1}{2}\tilde{\Omega}_{a}^2 \,\Big |_{d} &=&
  -2 e^{a N}\partial_{N}\left(e_{a}{}^{M}\partial_{M}d\right)
  -2\partial_{M}e_{a}{}^{M}\,e^{aN}\partial_{N}d+2e_{a}{}^{M}\partial_{M}d\,e^{a N}\partial_{N}d \\ \nonumber
  &=& -2 e^{aN}e_{a}{}^{M}\partial_{M}\partial_{N}d-2\partial_{N}\left(e^{aN} e_{a}{}^{M}\right)
  \partial_{M}d+2e_{a}{}^{M}\,e^{aN}\partial_{M}d\,\partial_{N}d \; .
\end{eqnarray}
With the expression for ${\cal H}^{MN}$ from (\ref{Hroot}) this reduces to
\be\label{RHd}
 e_{a}\tilde{\Omega}^{a}+\frac{1}{2}\tilde{\Omega}_{a}^2 \,\Big |_{d} \ = \
 {\cal H}^{MN}\partial_{M}\partial_{N}d + \partial_{N}{\cal H}^{MN}\,\partial_{M}d
 -{\cal H}^{MN}\partial_{M}d\,\partial_{N}d\;,
\ee
where we used the strong constraint (\ref{ODDconstr}).

We turn next to the pure $e$-dependent terms which are more involved. The first two terms in
(\ref{Rredux}) yield
 \be\label{H1}
  e_{a}\tilde{\Omega}^{a} + \frac{1}{2}\tilde{\Omega}_{a}^2 \,\Big |_{e} \ = \
  \frac{1}{2} \left( \partial_M \partial_N (e_{a}{}^{N}e^{aM}) - \partial_N e^{aM} \partial_M  e_{a}{}^{N} \right)
  \ = \
  -\frac{1}{4} \partial_M \partial_N   {\cal H}^{MN} - \frac{1}{2} \partial_N e^{aM} \partial_M  e_{a}{}^{N} \, .
\ee
In order to compute the third term in (\ref{Rredux}) we start from
 \be\label{start}
  \Omega_{ab\bar{c}}  \ = \
  2 f_{[ab] \bar{c}} +  f_{\bar{c} [ab]}
  \ = \ 2 \big( e_{[a} e_{b]}{}^{M}\big) e_{\bar{c}M} + \big(e_{\bar{c}} e_{a}{}^{M}\big) e_{bM} \;,
 \ee
where (\ref{simplyf}) implies automatic antisymmetry in $a,b$ in the last term.
Using that for arbitrary functions $X$ and $Y$ the strong constraint (\ref{flatconstr}) 
implies together with (\ref{Hroot}) 
\be
 e_{a}X\, e^{a}Y \ = \ -\frac{1}{2}{\cal H}^{MN}\partial_{M}X\,\partial_{N}Y\;, \qquad
 e_{\bar{a}}X\,e^{\bar{a}}Y \ = \  \frac{1}{2}{\cal H}^{MN}\partial_{M}X\,\partial_{N}X\;,
\ee
the square of (\ref{start}) reads
 \begin{eqnarray}\label{H2}
\Omega_{ab\bar{c}}{}^2 &=&
- \frac{1}{2} {\cal H}^{KL} ({\cal H}_{MN} + \eta_{MN}) \partial_K e_{b}{}^{M} \partial_L e^{bN}
 - ({\cal H}_{MN} + \eta_{MN}) e_{b} e_{a}{}^{M}\, e^{a} e^{bN}
\\ \nonumber
&& - \frac{1}{4} {\cal H}^{KL} ({\cal H}_{MN} - \eta_{MN}) \partial_K e_{b}{}^{M} \partial_L e^{bN} + ({\cal H}^{NK} - \eta^{NK}) ({\cal H}_{ML} + \eta_{ML}) \partial_K e_{b}{}^{M} \partial^L e^{b}{}_{N} \, .
\end{eqnarray}
We next compute with (\ref{antisol})
 \be
  \Omega_{[abc]}{}^2 \ = \ 9  f_{[abc]}{}^2 \ =  \
  3 ( f_{abc} + 2 f_{cab} ) f^{abc} ,
\ee
where (\ref{simplyf}) has been used. This yields
 \be\label{H3}
 \begin{split}
 \Omega_{[abc]}{}^2 \ &= \ 
 3 \left[ \frac{1}{4}{\cal H}^{KL} ({\cal H}_{MN} - \eta_{MN}) \partial_K e_{b}{}^M \, \partial_L e^{bN} 
 +2\big(e_{c}e_{a}{}^{M}\big)e_{bM}\big(e^{a}e^{bN}\big)e^{c}{}_{N}\right] \\
 \ &= \ \frac{3}{4} \big[{\cal H}_{KL} ({\cal H}_{MN} - \eta_{MN})  -2 ({\cal H}_{ML}
 - \eta_{ML}) ({\cal H}_{NK} - \eta_{NK}) \big] \partial^K e_{b}{}^M \, \partial^L e^{bN} \, ,
\end{split}
\ee
where we used in the second line (\ref{simplyf}).
In total, the third and the fourth term of ${\cal R}$ in (\ref{Rredux}) combine as follows:
\begin{eqnarray}\nonumber
-\frac{1}{4}\Omega_{ab\bar{c}}{}^{2}
   -\frac{1}{12}\Omega_{[abc]}{}^{2}  
    &=&  \frac{1}{8} {\cal H}^{KL} ({\cal H}_{MN} + \eta_{MN}) \partial_K e_{b}{}^{M} \partial_L e^{bN} 
    + \frac{1}{4} ({\cal H}_{MN} + \eta_{MN}) e_{b} e_{a}{}^{M}\, e^{a} e^{bN}
    \\ \label{Hinterm}
      && - \frac{1}{8}({\cal H}^{NK} - \eta^{NK}) ({\cal H}_{ML} + 3\eta_{ML}) \partial_K e_{b}{}^{M} \partial^L e^{b}{}_{N} \, .
\end{eqnarray}

Adding (\ref{H1}) and (\ref{Hinterm}) one obtains after some work
 \begin{eqnarray}
{\cal R} \,\Big |_{\cal H} 
   &=& -\frac{1}{4} \partial_M \partial_N {\cal H}^{MN} + \frac{1}{32} {\cal H}^{KL} \partial_K  {\cal H}_{MN}  \partial_L {\cal H}^{MN}  -\frac{1}{8} {\cal H}^{ML} \partial_K {\cal H}_{MN} \partial_L {\cal H}^{NK} \, .
\end{eqnarray}
In combination with (\ref{RHd}) we obtain in total 
 \begin{eqnarray}
{\cal R}   &=& {\cal H}^{MN}\partial_{M}\partial_{N}d + \partial_{N}{\cal H}^{MN}\,\partial_{M}d
   -{\cal H}^{MN}\partial_{M}d\,\partial_{N}d  \\[0.7ex] \nonumber
&& -\frac{1}{4} \partial_M \partial_N {\cal H}^{MN} + \frac{1}{32} {\cal H}^{KL} \partial_K  {\cal H}^{MN}  \partial_L {\cal H}_{MN}  -\frac{1}{8} {\cal H}^{ML}\partial_L {\cal H}^{NK} \partial_K {\cal H}_{MN}  \, .
\end{eqnarray}
This coincides with the curvature scalar ${\cal R}({\cal H},d)$ constructed in \cite{Hohm:2010pp},
up to the same irrelevant overall factor of 4 encountered above,  
and thus we have established independently the equivalence of the two action principles.

\subsection{${\cal H}$-compatible Christoffel connections} 
So far we have exclusively dealt with covariant derivatives acting on objects with `flat' or tangent 
space indices. 
Given these spin connection-type objects there is a canonical way to associate corresponding 
Christoffel-type connections, via the so-called vielbein postulate. 
Here we investigate the properties of these Christoffel symbols. 
Very recently, an interesting paper appeared that deals with the 
geometrical foundation of the ${\cal H}$-formulation and introduces similar connections \cite{Jeon:2010rw},
which are related but not identical to those discussed here.

We start by defining Christoffel symbols from the $GL(D)\times GL(D)$ connections by requiring 
 \be\label{Gammacov}
  \nabla_{M}V_{N} \ := \  \partial_{M}V_{N}-\Gamma_{MN}{}^{K}V_{K} \ \equiv \ 
  e_{M}{}^{A}e_{N}{}^{B}\nabla_{A}V_{B}\;,
 \ee
and analogously for higher tensors.  
This is satisfied if the following `vielbein postulate' holds 
 \be\label{vielpost}
  \partial_{M}e_{N}{}^{A}-\omega_{MB}{}^{A}e_{N}{}^{B}-\Gamma_{MN}{}^{K}e_{K}{}^{A} \ = \ 0\;,
 \ee
which is the usual condition that the vielbein is covariantly constant with respect to the
tangent space and Christoffel connections. This condition determines the Christoffel symbols 
in terms of (derivatives of) $e$ and $\omega$. Thus, $\Gamma$ is uniquely determined 
by the physical fields whenever this holds for $\omega$. Moreover, as the \textit{conventional}
torsion for $\omega$ is non-zero, there is a non-zero antisymmetric part $\Gamma_{[MN]}{}^{K}$
proportional to this torsion. With this covariant derivative curvature tensors may be defined 
via $[\nabla_{M},\nabla_{N}]$, and the resulting objects will be equivalent (through the conversion 
of indices with the frame field) to the corresponding 
tensors defined via (\ref{gencomm}), and thus all the comments there readily apply in the 
present context. 

It is instructive, however, to inspect some properties of the covariant derivatives based on 
$\Gamma$ in more detail. First,
from (\ref{vielpost}) we infer the transformation rule of $\Gamma$ under $\xi^{M}$ transformations, 
 \be
  \delta_{\xi}\Gamma_{MN}{}^{K} \ = \  \widehat{\cal L}_{\xi}\Gamma_{MN}{}^{K}
  +\partial_{M}\partial_{N}\xi^{K}-\partial_{M}\partial^{K}\xi_{N}\;.
 \ee
One may easily verify that this is the right transformation rule that makes the first expression in
(\ref{Gammacov}) a covariant derivative. The first inhomogeneous term is the standard one 
appearing for the Christoffel symbols in Riemannian geometry, while the second one is novel and
due to the generalized Lie derivative.  This new contribution also shows that 
$\delta_{\xi}\Gamma_{[MN]}{}^{K}\neq 0$ and thus that the connection is necessarily torsionful. 

An important consequence of the defining relation (\ref{Gammacov}) is that $\eta_{MN}$
is covariantly constant,
 \be
  \nabla_{M}\eta_{NK} \ = \ e_{M}{}^{A}e_{N}{}^{B}e_{K}{}^{C}\nabla_{A}{\cal G}_{BC} \ = \ 0\;,
 \ee
which follows from the definition (\ref{flatmetric}) and the metricity condition (\ref{covconst}) for the tangent space metric. 
Explicitly, this implies for the Christoffel symbols with (\ref{Gammacov}) 
 \be\label{Gammsym}
  \nabla_{M}\eta_{NK} \ = \ \partial_{M}\eta_{NK} -\Gamma_{MN}{}^{L}\eta_{LK}
  -\Gamma_{MK}{}^{L}\eta_{NL} \ = \  0 \qquad\Rightarrow\qquad
  \Gamma_{M(NK)} \ = \ 0\;.
 \ee
Moreover, since the `flattened' components of ${\cal H}$ are given by the components of ${\cal G}^{AB}$,
up to sign differences that account for the different signatures, c.f.~eq.~(\ref{Hflat}), the 
metricity of the tangent space metric implies also 
  \be
   \nabla_{M}{\cal H}_{NK} \ = \ e_{M}{}^{A}e_{N}{}^{B}e_{K}{}^{C}\nabla_{A}{\cal H}_{BC} \ = \ 0\;,
 \ee
and therefore $\Gamma$ is an ${\cal H}$-compatible connection. In other words, in this formalism there are two covariantly constant metrics, ${\cal H}$ and $\eta$. 

Another important property of the Christoffel symbols follows from the vielbein postulate (\ref{vielpost}),
 \be\label{Gammator}
  \Gamma_{[MNK]} \ = \ -f_{[MNK]} -\omega_{[MNK]} \ = \ 0\;,
 \ee
where all indices have been converted into world indices, and the last equation follows from (\ref{antisol}).
As the latter equation was a direct consequence of the (generalized) torsion constraint, 
eq.~(\ref{Gammator}) can be seen as the analogue of the usual torsion constraint 
$\Gamma_{[MN]K}=0$. The properties (\ref{Gammsym}) and (\ref{Gammator}) imply that in 
the generalized Lie derivatives the partial derivatives can be replaced by covariant derivatives 
\cite{Jeon:2010rw},
 \be
 \begin{split}
  \widehat{\cal L}_{\xi}V_{M} \ &= \ \xi^{N}\partial_{N}V_{M}+\left(\partial_{M}\xi^{N}-\partial^{N}\xi_{M}\right)
  V_{N} \\
   \ &= \ \xi^{N}\nabla_{N}V_{M}+\left(\nabla_{M}\xi^{N}-\nabla^{N}\xi_{M}\right)V_{N}
   +\xi^{N}V^{K}\left(\Gamma_{NMK}-\Gamma_{MNK}-\Gamma_{KMN}\right) \\
   \ &= \ \xi^{N}\nabla_{N}V_{M}+\left(\nabla_{M}\xi^{N}-\nabla^{N}\xi_{M}\right)V_{N}\;,
  \end{split}
  \ee
and similarly on arbitrary higher tensors. In conventional Riemannian geometry the usual Lie derivative
has the analogous property by virtue of the usual torsion constraint.

The frame formalism carries only connections with respect to $GL(D)\times GL(D)$, and so 
this `factorization' should also be visible in the Christoffel symbols of the generalized metric formulation.
To see this, we note that due to the constraint (\ref{Hconst}) on ${\cal H}$, the matrices 
  \be
   {\cal P}_{M}{}^{N} \ = \ \frac{1}{2}\left(\delta_{M}{}^{N}-{\cal H}_{M}{}^{N}\right)\;, \qquad
   \bar{\cal P}_{M}{}^{N} \ = \ \frac{1}{2}\left(\delta_{M}{}^{N}+{\cal H}_{M}{}^{N}\right)
 \ee
are projectors satisfying ${\cal P}^2={\cal P}$ and $\bar{\cal P}^2=\bar{\cal P}$.
With the expression (\ref{Hroot}) for ${\cal H}$ in terms of the frame fields, this simply reduces
to \cite{Hohm:2010pp}
 \be
   {\cal P}_{M}{}^{N} \ = \ e_{aM}e^{aN}\;, \qquad
   \bar{\cal P}_{M}{}^{N} \ = \  e_{\bar{a}M}e^{\bar{a}N}\;.
 \ee
Therefore, $\bar{\cal P}$ and ${\cal P}$ project onto the subspaces that are invariant under the left
$GL(D)$ or the right $GL(D)$, respectively. More precisely, given a vector $V_{M}$ that is projected 
to the left (`unbarred') subspace, i.e., invariant under the right $GL(D)$, we find indeed 
 \be
   V_{M} \ = \ {\cal P}_{M}{}^{N}V_{N} \ = \ e_{aM}e^{aN}V_{N} \qquad \Rightarrow \qquad
   V_{\bar{a}} \ \equiv \ e_{\bar{a}}{}^{M}V_{M} \ = \ 0\;,
 \ee
and thus only $V_{a}$ is non-zero. Here, in the last step, (\ref{Goff}) has been used.      
Analogously, the frame components of a vector with $V_{M}=\bar{\cal P}_{M}{}^{N}V_{N}$ 
satisfy $V_{a}=0$. 

It is now straightforward to see that the covariant derivative is compatible with these projections.
In fact, since ${\cal H}$ and $\eta$ are covariantly constant, we find
 \be\label{Pconst}
  V_{M} \ = \ {\cal P}_{M}{}^{N}V_{N} \qquad\Rightarrow \qquad
  \nabla_{M}V_{N} \ = \ {\cal P}_{N}{}^{K}\nabla_{M}V_{K}\;,
 \ee
and similarly for $\bar{\cal P}$. This is the analogue of the fact, which is manifest in the 
frame formalism, that the $GL(D)\times GL(D)$ covariant derivatives preserve the 
barred-unbarred index structure.   

The Christoffel symbols discussed here are closely related to those introduced in \cite{Jeon:2010rw}. 
First, the property that the 
covariant derivatives preserve the left- and right-invariant subspaces as in (\ref{Pconst}) 
is one of the requirements that determines their $\Gamma$. Second, the Christoffel symbols are
further constrained in \cite{Jeon:2010rw} by requiring 
(\ref{Gammsym}) and (\ref{Gammator}). 
The details of the connections 
differ, however, in that their covariant derivatives do not transform covariantly, but only in certain
combinations and projections, while Siegel's connections -- and thereby the Christoffel 
symbols determined by (\ref{vielpost}) -- properly transforms as connections, at the cost of 
introducing components that are not determined in terms of the physical fields.

\section{Summary and Outlook}\setcounter{equation}{0}
Recent results on double field theory have given an elegant 
`$O(D,D)$ covariantisation' of the conventional low-energy space-time action (\ref{original}) 
of closed string theory 
by virtue of introducing extra coordinates. The resulting actions, written in terms of ${\cal E}_{ij}$ 
or ${\cal H}^{MN}$, take a remarkably simple form
and feature besides the global $O(D,D)$ T-duality invariance a gauge symmetry that 
unifies the usual diffeomorphisms with the 2-form gauge symmetry.  So far, however, a deeper 
understanding of the geometrical structure of this theory, adopting the role that 
Riemannian geometry plays in Einstein's theory, was lacking. In this paper we have shown that 
the duality-covariant geometrical  formalism developed by Siegel already some time ago in 
\cite{Siegel:1993th} provides, at least to some extent, such a framework in terms of frame fields, connections 
and curvatures for the gauge group $GL(D)\times GL(D)$. For the convenience of the reader 
we summarize here the main differences to ordinary Riemannian geometry. 

First of all, a central object is the $O(D,D)$ invariant metric $\eta$ which is a constant 
`world tensor' with two upper or two lower indices. In Riemannian geometry such an object  
would not be well-defined, but here the constancy of $\eta$ has a gauge invariant 
meaning due to the modified form of the gauge transformations, governed by the `generalized 
Lie derivatives' (\ref{genLie}). In contrast to the `world' metric $\eta^{MN}$, 
the `tangent space metric' ${\cal G}_{AB}$ is space-time dependent, and thus we have the opposite of the usual 
situation. It is instructive to compare this with a reformulation of conventional Riemannian geometry that resembles 
the formalism presented here in that there is an enlarged group of frame transformations, the general 
linear group $GL(D)$ rather than the Lorentz group,  and a space-time dependent tangent space metric 
$g_{ab}$ that enters together with the vielbein $e_{a}{}^{m}$ as an independent field (see sec.~IX.A.2 in \cite{Siegel:1999ew}). Imposing a metricity condition and the usual torsion constraint,
 \be\label{concleq}
  \nabla_{a}g_{bc} \ = \ 0\;, \qquad 
  T_{ab}{}^{c} \ = \ -2e_{a}{}^{m}e_{b}{}^{n}\nabla_{[m}e_{n]}{}^{c} \ = \ 0\;,
 \ee
allows one to solve for the connections $\omega_{abc}$ in terms of derivatives of $e_{a}{}^{m}$
and $g_{ab}$. The local $GL(D)$ symmetry can then be fixed by setting either $e_{a}{}^{m}=\delta_{a}{}^{m}$,
in which case $g_{ab}$ can be identified with the usual metric and the $\omega_{abc}$ reduce
to the Christoffel symbols $\Gamma_{abc}$, or one can set $g_{ab}=\delta_{ab}$, in which case
$e_{a}{}^{m}$ carries the physical degrees of freedom and $\omega_{abc}$ reduces to the 
usual spin connection.   
This formalism differs, however, from the present frame formalism, at least in the 
form discussed in this paper, in several respects. For instance, here it is not the tangent space metric 
${\cal G}_{AB}$ that is introduced as an independent object but rather the constant 
$O(D,D)$ invariant metric $\eta^{MN}$, while ${\cal G}_{AB}$ is defined in terms of $\eta_{MN}$ 
by use of the frame fields.  Moreover, the torsion constraint is modified as compared to (\ref{concleq}). 

Perhaps the most important difference to Riemannian geometry is the novel gauge symmetry
parametrized by $\xi^{M}$, whose
algebra is governed by the C-bracket rather than the Lie bracket of the usual diffeomorphisms. This  
has a number of consequences. 
Most importantly, due to the modified torsion constraint, 
the Riemann-like tensor 
defined through the commutator of covariant derivatives is generally not covariant under 
frame rotations.
Following \cite{Siegel:1993th} this can be repaired `by hand', but is should be 
stressed that the resulting tensor, which is fully covariant, 
is not in all components independent on the undetermined connections.  
The resulting Ricci-like tensor and scalar curvature are, however, fully expressible in terms of the 
physical fields, and are equivalent to the field equations and Lagrangian of double field theory, 
respectively. 

It is natural to anticipate that a yet better understanding of the geometrical structure is possible, 
perhaps adopting and extending ideas from `generalized geometry' \cite{Tcourant,Hitchin,Gualtieri}, 
in which, for instance, a fully covariant curvature tensor may emerge more directly. 
Such an understanding could be useful not only for the double field theory currently discussed, 
but also for further generalizations, say, to type II string theory. Finally, we hope that the present 
investigations might shed some light on the possibility of the ultimate goal of this research program,
namely to construct a `truly doubled field theory' in which the strong constraint  (\ref{ODDconstr})
is relaxed in such a way that the fields may depend non-trivially on both momentum and winding coordinates 
even locally.

\subsection*{Acknowledgments}
We are happy to acknowledge helpful discussions with Ashoke Sen and especially Barton Zwiebach.
We have also benefitted from discussions with Chris Hull at initial stages of this project and later correspondence. 
This work is supported by the U.S. Department of Energy (DoE) under the cooperative
research agreement DE-FG02-05ER41360. The work of OH is supported by the DFG -- The German Science Foundation. The work of SK is supported in part by a Samsung Scholarship.

\appendix

\section{Computational details on the curvature tensor}
\setcounter{equation}{0}

\subsection{Fully contracted curvature tensor without constraints}
In this appendix we prove the equation (\ref{offscalar}) which holds before imposing any constraints.
Thus, we have to use the form of the curvature tensor in (\ref{offcovtensor}), which was fully
covariant without using constraints. More explicitly, this reads
 \begin{eqnarray}\label{calRfull}
  {\cal R}_{ABCD} &=&  \frac{1}{2}\big[R_{ABCD}+R_{CDAB}\big]\\ \nonumber
  &&-\frac{1}{4}\big[\omega_{ECD}\,
  \omega^{E}{}_{AB}+\omega_{EAB}\,\omega^{E}{}_{CD}\big]
  -\frac{1}{4}\big[\omega_{ECD}\,e^{E}{\cal G}_{AB}+\omega_{EAB}\,e^{E}{\cal G}_{CD}\big]\;,
\end{eqnarray}
from which we derive
\be\label{calRfull2}
{\cal R}_{AB}{}^{AB} \ = \ {\cal G}^{AC}{\cal G}^{BD}{\cal R}_{ABCD}
\ = \ {\cal G}^{AC} R_{ABC}{}^{B} - \frac{1}{2} \omega_{E}{}^{AB} \,e^{E}{\cal G}_{AB} -\frac{1}{2}\omega_{E}{}^{AB}\,\omega^{E}{}_{AB} \, .
\ee
We rewrite now the first term on the right-hand side as follows
 \begin{eqnarray}\label{step9721}
R_{AB}{}^{AB} &=&  {\cal G}^{AC} \Big( e_{A}\omega_{BC}{}^{B}-e_{B}\omega_{AC}{}^{B}
  +\omega_{AC}{}^{E}\omega_{BE}{}^{B}-\omega_{BC}{}^{E}\omega_{AE}{}^{B}
  -\Omega_{AB}{}^{E}\omega_{EC}{}^{B} \Big) \\
  &=&  {\cal G}^{AC} \Big( \nabla_A \omega_{BC}{}^{B} - \nabla_B \omega_{AC}{}^{B}  -\omega_{BE}{}^B \omega_{AC}{}^E  \Big) + \Big(-\Omega_{AB}{}^{E} + \omega_{BA}{}^E  \Big) \omega_{E}{}^{AB} \nonumber \,.
\end{eqnarray}
In the first bracket we next move the metrics inside the covariant derivatives, where we have
to remember that here the metric is not assumed to be covariantly constant,
  \begin{eqnarray}
{\cal G}^{AC} \Big( \nabla_A \omega_{BC}{}^{B} - \nabla_B \omega_{AC}{}^{B}  -\omega_{BE}{}^B \omega_{AC}{}^E  \Big)  &=&  \nabla_A \omega_{B}{}^{AB} - \omega_{BC}{}^{B} \nabla_A {\cal G}^{AC}
- \nabla_B \omega_{A}{}^{AB}\\
 &&+ \omega_{AC}{}^{B}  \nabla_B {\cal G}^{AC} - \omega_{BE}{}^B
 \omega_{A}{}^{AE} \nonumber \, .
\end{eqnarray}
In order to eliminate the last term proportional to $\omega^2$, we use
 \begin{eqnarray}
  \nabla_{A}\nabla_{B}{\cal G}^{AB} &=& e_{A}e_{B}{\cal G}^{AB}-\nabla_{A}\omega_{B}{}^{BA}
  -\nabla_{A}\omega_{B}{}^{AB}\\ \nonumber
  &&-\omega_{AC}{}^{A}\,\nabla_{B}{\cal G}^{CB}-\omega_{BE}{}^{B}\,\omega_{A}{}^{AE}
  -\omega_{BE}{}^{B}\,\omega_{A}{}^{EA}\;,
 \end{eqnarray}
which is straightforward to verify. This leads to
 \begin{eqnarray}
{\cal G}^{AC} \Big( \nabla_A \omega_{BC}{}^{B} - \nabla_B \omega_{AC}{}^{B}  -\omega_{BE}{}^B \omega_{AC}{}^E  \Big)
&=& 2 \nabla_A \omega_{B}{}^{AB} + \nabla_A \nabla_B  {\cal G}^{AB} - e_A e_B  {\cal G}^{AB}
\\ \nonumber
&&+  \omega_{AC}{}^{B}  \nabla_B {\cal G}^{AC} + \omega_{BE}{}^B   \omega_{A}{}^{EA} \, .
\end{eqnarray}
We use next the definition $\tilde{T}_A = \omega_{BA}{}^B  + f_{BA}{}^B - 2e_A d$ in order to
eliminate $\omega_{BA}{}^B$ in the first and last term by $\tilde{T}$ and $f$. Moreover,
we use $2f_{A(BC)} = e_A {\cal G}_{BC}$ in order to rewrite the third term in terms of $f$.
This leads after a short computation to
\be
\begin{split}
 {\cal G}^{AC} \Big( &\nabla_A \omega_{B\,C}{}^{B} - \nabla_B \omega_{AC}{}^{B}  -\omega_{BE}{}^B \omega_{AC}{}^E  \Big) \\[0.5ex]
 \ = \  &2 \nabla_A \tilde{T}^A + 2 e_A f_{B}{}^{[AB]}  + 2 \omega_{AE}{}^{A}   f_{B}{}^{E B} 
 + \nabla_A \nabla_B  {\cal G}^{AB}
 +  \omega_{AC}{}^{B}  \nabla_B {\cal G}^{AC} \\[0.5ex]
 &+ \tilde{T}_A{}^2 + f_{BE}{}^B   f_{A}{}^{EA}  
   - 2  \tilde{T}_A f_{B }{}^{A B} - 4 f_{BE}{}^B e^E d + 4\big(\nabla_A e^A d + \tilde{T}_A e^A d\big)
 \, ,
\end{split}
\ee
where we have used the constraint (\ref{flatconstr}).
In here, the terms in brackets in the last line vanish by (\ref{Ttilde2}).
By finally eliminating $\omega_{AE}{}^A$ in the third term above in terms of $\tilde{T}$ and $f$, we arrive at
 \be
 \begin{split}
{\cal G}^{AC} \Big( &\nabla_A \omega_{BC}{}^{B} - \nabla_B \omega_{AC}{}^{B}  -\omega_{BE}{}^B \omega_{AC}{}^E  \Big) \\
 \ = \  &2 \nabla_A \tilde{T}^A +  \tilde{T}_A{}^2 + \nabla_A \nabla_B  {\cal G}^{AB}  + 2 e_B  f_{A}{}^{[AB]} - f_{BE}{}^B   f_{A}{}^{EA} +  \omega_{AC}{}^{B}  \nabla_B {\cal G}^{AC}
\, .
\end{split}
\ee
Using this in (\ref{step9721}) we get
 \begin{eqnarray}
  R_{AB}{}^{AB} &=& 2 \nabla_A \tilde{T}^A +  \tilde{T}_A{}^2 + \nabla_A \nabla_B  {\cal G}^{AB}  \\ \nonumber
  &&+  2 e_B  f_{A}{}^{[AB]} - f_{BE}{}^B   f_{A}{}^{EA} +  \omega_{AC}{}^{B}  \nabla_B {\cal G}^{AC} 
  + \big(-\Omega_{AB}{}^{E} + \omega_{BA}{}^E  \big) \omega_{E}{}^{AB} \, .
\end{eqnarray}

To proceed with the computation of the full ${\cal R}_{AB}{}^{AB}$ in (\ref{calRfull2}) we
define
\begin{eqnarray}
\Delta{\cal R} &\equiv& 2 e_B  f_{A}{}^{[AB]} - f_{BE}{}^B   f_{A}{}^{EA} +  \omega_{AC}{}^{B}  \nabla_B {\cal G}^{AC} + (-\Omega_{AB}{}^{E} + \omega_{BA}{}^E  ) \omega_{E}{}^{AB} \\ \nonumber && - \frac{1}{2} \omega_{E}{}^{AB} \,e^{E}{\cal G}_{AB} -\frac{1}{2}\omega_{E}{}^{AB}\,\omega^{E}{}_{AB} \, ,
\end{eqnarray}
such that
\begin{eqnarray}
{\cal R}_{AB}{}^{AB} \ = \ 2 \nabla_A \tilde{T}^A +  \tilde{T}_A{}^2 + \nabla_A \nabla_B  {\cal G}^{AB} + \Delta{\cal R} \, .
\end{eqnarray}
To evaluate $\Delta{\cal R}$, we need the following identities:
\begin{eqnarray}
2 e_B  f_{A}{}^{[AB]} &=&  f_{B E}{}^B   f_{A}{}^{EA} -  f_{AB}{}^E   f_{E}{}^{AB}   \, , \\
\nabla_A {\cal G}_{BC}  &=& e_A {\cal G}_{BC} + \omega_{ABC} + \omega_{AC B} \ = \ 2 f_{A (BC) } + 2 \omega_{A (BC)}\;. 
\end{eqnarray}
Using the definition $\Omega_{A  B}{}^{C} = 2 f_{[AB]}{}^C +  f^C{}_{[AB]}$, we can rewrite $\Delta{\cal R}$
in terms of only $f$ and $\omega$ as follows:
\be
 \Delta{\cal R} \ = \ - \left( f_{ABE}   f^{EAB}  + 2\, \omega_{ABE}   f^{EAB} + \omega_{ABE}
 \omega^{EAB}\right) -\frac{1}{2} \left( 2 \, \omega^{EAB} f_{EAB} + \omega^{EAB} \omega_{EAB}  \right)  \, .
\ee
By adding $-\frac{1}{2} f^{EAB} f_{EAB}$, which vanishes due to the strong constraint, $\Delta{\cal R}$
can be written as
\be
 \Delta{\cal R} \ = \ -  K^{EAB} K_{ABE}   -\frac{1}{2}  K^{EAB} K_{EAB}
 \ = \ - \frac{1}{2} K^{EAB} \big(K_{ABE} + K_{BEA} + K_{EAB} \big) \, ,
\ee
where we introduced
 \be
  K_{A  B}{}^{C} \ = \ f_{A  B}{}^{C} + \omega_{A  B}{}^{C}\;.
 \ee
We can further rewrite $\Delta{\cal R}$ according to
\be
 \Delta{\cal R} \ = \ -  \frac{3}{2} K_{[ABC]}{}^2 - \frac{3}{2} K_{(ABC)}{}^2 \, .
\ee
Next, we use the following two identities
\be
 {\cal T}_{[ABC]}  \ = \ 3 K_{[ABC]}  \;, \qquad
 K_{(ABC)} \ = \ \frac{1}{2} \nabla_{(A} {\cal G}_{BC)} \, ,
\ee
which can be easily confirmed,
to obtain
\be
 \Delta{\cal R} \ = \ - \frac{1}{6} {\cal T}_{[ABC]}{}^2 - \frac{3}{8} \nabla_{(A} {\cal G}_{BC)}{}^2 \;.
\ee
This finally leads to
\be
{\cal R}_{AB}{}^{AB} \ = \ 2 \nabla_A \tilde{T}^A +  \tilde{T}_A{}^2 + \nabla_A \nabla_B  {\cal G}^{AB} - \frac{1}{6} {\cal T}_{[ABC]}{}^2 - \frac{3}{8} \nabla_{(A} {\cal G}_{BC)}{}^2 \;,
\ee
as we wanted to show.

\subsection{Scalar curvature}
In this appendix we prove that the curvature scalar given by the first expression in (\ref{originalscalar}) can be written as (\ref{finalscalar}).
From the definition of the Riemann tensor \eqref{modified_riemann} we obtain
\be \label{ricci_scalar_expression}
{\cal R} \ = \ -\frac{1}{2}{\cal R}_{ab}{}^{ab} \ = \ 
-\frac{1}{2}R_{ab}{}^{ab} - \frac{1}{4} \omega_{Eab}\, \omega^{Eba} \, ,
\ee
where $R_{ab}{}^{ab}$ is given by
\begin{eqnarray}
R_{ab}{}^{ab} &=& {\cal  G}^{ac} \bigg[ e_a \omega_{bc}{}^b - e_b \omega_{ac}{}^{b} + \omega_{ac}{}^e \omega_{be}{}^b - \omega_{bc}{}^e \omega_{ae}{}^b - \Omega_{ab}{}^E \omega_{Ec}{}^b \bigg] \\ [1.0ex] \nonumber
&=& - e_a \tilde{\Omega}^a + (e_a {\cal G}^{ac}) \tilde{\Omega}_c - e_b ({\cal G}^{ac} \omega_{ac}{}^b)  + (e_b {\cal G}^{ac}) \omega_{ac}{}^b + \omega_a{}^{ae} \omega_{be}{}^b - \omega_b{}^{ae} \omega_{ae}{}^b - \Omega_{ab}{}^E \omega_{E}{}^{ab} \, .
\end{eqnarray}
Using
\be
 \omega_{a}{}^{ab} \ = \ \tilde{\Omega}^b - {\cal G}^{bc} e^a {\cal G}_{ac}
 \ = \ \tilde{\Omega}^b + e_a {\cal G}^{ab} \, ,
\ee
which follows from (\ref{tracepart}) and (\ref{metrsol}), a short computation shows that this can
be rewritten as
\begin{eqnarray}
R_{ab}{}^{ab}
&=& - 2 e_a \tilde{\Omega}^a - \tilde{\Omega}_a{}^2  - e_a e_b {\cal G}^{ab} + \Big[ \omega_{bca} \omega^{acb} - \Omega_{ab}{}^E \omega_{E}{}^{ab} \Big] \, .
\end{eqnarray}
Then ${\cal R}$ is according to  \eqref{ricci_scalar_expression} given by
\be\label{step0}
{\cal R} \ = \ -\frac{1}{2}{\cal R}_{ab}{}^{ab} = e_a \tilde{\Omega}^a +\frac{1}{2} \tilde{\Omega}_a{}^2  
+\frac{1}{2} e_a e_b {\cal G}^{ab} -\frac{1}{2} \Big[ \omega_{bca} \omega^{acb} - \Omega_{ab}{}^E \omega_{E}{}^{ab} + \frac{1}{2} \omega_{Eab} \omega^{Eba} \Big] \, .
\ee
We next evaluate the terms in the square bracket. Using the torsion constraint (\ref{constr}) we find
\begin{eqnarray} \label{detail_216}
\omega_{bca} \omega^{acb} - \Omega_{ab}{}^E \omega_{E}{}^{ab} + \frac{1}{2} \omega_{Eab} \omega^{Eba}  &=& \omega_{abc} \omega^{cab} + \frac{1}{2} \omega_{cab} \omega^{cab}  + \frac{1}{2} \omega_{\bar{c}ab} \omega^{\bar{c}ab}  \\ \nonumber
&=&\frac{1}{2} \big[ \omega_{abc} +  \omega_{bca} + \omega_{cab} \big]  \omega^{cab} + \frac{1}{2}  \omega_{\bar{c}ab} \omega^{\bar{c}ab}  \, ,
\end{eqnarray}
where in the second equation $\omega_{bca}\omega^{cab}=\omega_{abc}\omega^{cab}$ has been
used.
The terms in the square bracket in the second line can be written as
\begin{eqnarray}
\frac{1}{2} \big[ \omega_{abc} +  \omega_{bca} + \omega_{cab} \big]  \omega^{cab} &=& \frac{1}{4} \big[ \omega_{abc} +  \omega_{bca} +\omega_{cab} -  \omega_{acb} -  \omega_{bac} -  \omega_{cba} \big] \omega^{cab}  \\ \nonumber  &&+ \frac{1}{4} \big(\omega_{cab} + \omega_{cba} \big) \omega^{cab}   +   \frac{1}{4} \big( \omega_{abc} +  \omega_{bca} +\omega_{acb} +  \omega_{bac} \big) \omega^{cab}  \, .
\end{eqnarray}
The terms in the each bracket give the following contributions, respectively:
\begin{eqnarray}\label{fEq1}
 \frac{3}{2} \omega_{[abc]} \omega^{[abc]}   &=& \frac{1}{6} \Omega_{[abc]}{}^2 
 \ = \ \frac{3}{2} f_{[abc]} f^{[abc]} \, ,
 \\ \label{fEq2}   \frac{1}{2} \omega_{c(ab)}  \omega^{c(ab)} &=&
  -  \frac{1}{8} e^a {\cal G}^{bc} e_a {\cal G}_{bc}  
  \ = \   \frac{1}{2} f_{c(ab)}  f^{c(ab)}\, , \\ \label{fEq3}
     \frac{1}{4} \big( \omega_{abc} +  \omega_{bca} +\omega_{acb} +  \omega_{bac} \big) \omega^{cab} &=&
     \omega_{a(bc)} \omega^{b(ac)}  \ = \ -  \frac{1}{4} e^a {\cal G}^{bc} e_b {\cal G}_{ac} 
     \ = \  f_{a(bc)} f^{b(ac)} \, .
\end{eqnarray}
Thus, \eqref{detail_216} can finally be written as
\be\label{step}
\omega_{bca} \omega^{acb} - \Omega_{ab}{}^E \omega_{E}{}^{ab} + \frac{1}{2} \omega_{Eab} \omega^{Eba}   \ = \ \frac{1}{6} \Omega_{[abc]}{}^2 -  \frac{1}{8} e^a {\cal G}^{bc} e_a {\cal G}_{bc} -  \frac{1}{4} e^a {\cal G}^{bc} e_b {\cal G}_{ac} + \frac{1}{2}  \omega_{\bar{c}ab} \omega^{\bar{c}ab} \, .
\ee
Using this in (\ref{step0}) together with $\omega_{\bar{c}ab}=-\Omega_{\bar{c}ab}$ we arrive at
 \be\label{betterR}
   {\cal R} \ = \ e_a \tilde{\Omega}^a + \frac{1}{2} \tilde{\Omega}_a{}^2  + \frac{1}{2} e_a e_b {\cal G}^{ab}
   -\frac{1}{12} \Omega_{[abc]}{}^2 +  \frac{1}{16} e^a {\cal G}^{bc} e_a {\cal G}_{bc} 
   + \frac{1}{8} e^a {\cal G}^{bc} e_b {\cal G}_{ac} - \frac{1}{4}  \Omega_{\bar{c}ab} \Omega^{\bar{c}ab} \;.
  \ee
With the expressions in terms of $f$ in (\ref{fEq1}), (\ref{fEq2}) and (\ref{fEq3}) this can also be written as 
 \be\label{Rsemifinal}
  {\cal R} \ = \ e_a \tilde{\Omega}^a + \frac{1}{2} \tilde{\Omega}_a{}^2  + \frac{1}{2} e_a e_b {\cal G}^{ab}
  -\frac{1}{4} \big(f_{abc} + f_{bca} + f_{cab}\big) f^{cab} - \frac{1}{4} \Omega_{\bar{c} ab} \Omega^{\bar{c} ab}\;.
 \ee 
  
The expressions for ${\cal R}$ given so far can be further simplified by using for the last term in (\ref{betterR})
the following relation    
 \be
  \Omega_{\bar{c}ab}  \ = \
    \Omega_{ab\bar{c}} +\frac{1}{2}e_{\bar{c}}{\cal G}_{ab}\;,
 \ee
which implies
\be\label{step2}
\frac{1}{2}\Omega_{\bar{c}ab}{}^2
 \ = \ \frac{1}{2}\Omega_{ab \bar{c}}{}^2 - \frac{1}{8} e^{\bar{c}} {\cal G}^{ab} e_{\bar{c}} \, {\cal G}_{ab}
 \ = \
 \frac{1}{2}\Omega_{ab \bar{c}}{}^2 + \frac{1}{8} e^{c} {\cal G}^{ab}\, e_{c} \, {\cal G}_{ab}  \, .
\ee
Here we used that the mixed term in the square can be brought to the form of the last term, using
$\Omega_{\bar{c}(ab)}=\tfrac{1}{2}e_{\bar{c}}{\cal G}_{ab}$.
In total,  using (\ref{step2}) in (\ref{betterR}) yields the following
expression for the scalar curvature:
\be\label{Ricciscalar}
{\cal R} \ = \ e_a \tilde{\Omega}^a + \frac{1}{2} \tilde{\Omega}_a{}^2  + \frac{1}{2} e_a e_b {\cal G}^{ab}   - \frac{1}{4} \Omega_{ a b \bar{c}} {}^2  -  \frac{1}{12} \Omega_{[abc]}{}^2 + \frac{1}{8} e^a {\cal G}^{bc} e_b {\cal G}_{ac}\;,
\ee
as we wanted to show.

\subsection{Scalar curvature in terms of $GL(D)\times GL(D)$ covariant derivatives}
Here we derive from the expression (\ref{betterR}) for the scalar curvature the 
manifestly $O(D,D)$ and $GL(D)\times GL(D)$ covariant form given in (\ref{explR}).  
We first focus on the dilaton-dependent terms which originate only from the first two 
terms in (\ref{betterR}), 
\begin{eqnarray}\label{step850}
{\cal R} \Big|_d &=&  e_a ( -2 e^{a} d ) + \frac{1}{2} \Big[ 2 {\cal G}^{ab} (\partial_M e_a{}^M) (-2 e_b d) +  (-2 e_a d)(-2 e^a d)  \Big] \\ \nonumber
&=&   -2 \Big( e_a e^a d + e^{2d} \partial_M (e_a{}^M e^{-2d}) e^a d + e_a d \, e^a d \Big) \\ \nonumber &=& -2 \Big(  e_a e^a d - \omega_{ba}{}^b e^a d + e_a d \, e^a d \Big) 
\ = \ -2 {\nabla}_a {\nabla}^a d -2 \nabla_a d \,\nabla^a d \, .
\end{eqnarray}
This reproduces the terms in (\ref{explR}) that are quadratic in the dilaton, if we use the strong 
constraint in the forms (\ref{flatconstr}) and (\ref{Ttilde2}). There are also apparent terms linear in 
the dilaton in (\ref{explR}) but these are actually artifacts of the dilaton dependence of some of 
the covariant derivatives. More precisely, terms in the first and last line of  (\ref{explR}) can be 
combined as follows 
 \be\label{step1004}
 \begin{split}
&- \frac{1}{2} \left(\nabla^a (e_{a}{}^M \bar{\nabla}^{\bar{b}} e_{\bar{b} M} ) - \bar{\nabla}^{\bar{a}} (e_{\bar{a}}{}^M \nabla^b e_{b M})  \right) - \left(\nabla^a d \, (e_{a}{}^M \bar{\nabla}^{\bar{b}} e_{\bar{b} M})  - \bar{\nabla}^{\bar{a}} d \, (e_{\bar{a}}{}^M {\nabla}^{b} e_{b M}  )\right) \\ 
&= \  - \frac{1}{2} \partial_M \left(e_a{}^M e^{a N} e^{\bar{b}} e_{\bar{b} N} - e_{\bar{a}}{}^M e^{\bar{a} N} e^{b} e_{b N} \right) \, .
\end{split}
\ee
This shows that this expression is independent of the dilaton and thus we have shown that 
(\ref{betterR}) reproduces the correct dilaton-dependent terms. For the following computation 
it is convenient to simplify the structure obtained in (\ref{step1004}) further, which yields 
\begin{eqnarray} \label{appFidentity}
  \frac{1}{2} \partial_M \left(e_{\bar{a}}{}^M e^{\bar{a} N} e^{b} e_{b N} - e_a{}^M e^{a N} e^{\bar{b}} e_{\bar{b} N} \right) &=& - \frac{1}{2} \partial_M \left( e_{b N} e^{b} (e_{\bar{a}}{}^M e^{\bar{a} N} ) - e_{\bar{b} N}  e^{\bar{b}} (e_a{}^M e^{a N}) \right) \\ \nonumber
  &=& \frac{1}{2}  \partial_M \left( e_{b N} e^{b} (e_a{}^M e^{a N}) +  e_{\bar{b} N}  e^{\bar{b}} (e_a{}^M e^{a N})\right)\\ \nonumber
  &=&  \frac{1}{2} \partial_M \partial_N \big(e_a{}^M e^{a N}\big) \, .
\end{eqnarray}

We next consider the terms in (\ref{betterR}) depending only on the frame field.  The first three terms in 
(\ref{betterR}) yield 
\begin{eqnarray}\label{step1843}
e_a \tilde{\Omega}^a + \frac{1}{2} \tilde{\Omega}_a{}^2  + \frac{1}{2} e_a e_b {\cal G}^{ab}
 \Big|_e  &=& 
e_a \big({\cal G}^{ab} \partial_N e_b{}^N\big) + \frac{1}{2} {\cal G}^{ab} \partial_M e_a{}^M \partial_N e_b{}^N +  \frac{1}{2} e_a e_b {\cal G}^{ab} \\ \nonumber 
&=& \frac{1}{2} \partial_M \big(e_a{}^M {\cal G}^{ab} \partial_N e_b{}^N\big) + \frac{1}{2} e_a \partial_N\big( e_b{}^N {\cal G}^{ab} \big) \\ \nonumber
&=& \frac{1}{2}  \partial_M  e^{a M} \partial_N e_a{}^N + \frac{1}{2} e^{a M}  \partial_M \partial_N e_a{}^N + \frac{1}{2}  e_a{}^M \partial_M \partial_Ne^{a N}  \\ \nonumber
&=& \frac{1}{2} \partial_M \partial_N \big(e^{a M} e_a{}^N\big)
 - \frac{1}{2} \partial_M e_a{}^N \partial_N e^{a M}  \, ,
\end{eqnarray}
which, combined with 
the relations \eqref{step1004} and (\ref{appFidentity}), implies
\begin{eqnarray}
e_a \tilde{\Omega}^a + \frac{1}{2} \tilde{\Omega}_a{}^2  + \frac{1}{2} e_a e_b {\cal G}^{ab} &=& - \frac{1}{2} \left(\nabla^a (e_{a}{}^M \bar{\nabla}^{\bar{b}} e_{\bar{b} M} ) - \bar{\nabla}^{\bar{a}} (e_{\bar{a}}{}^M \nabla^b e_{b M})  \right) \\ \nonumber
&&  - \left(\nabla^a d \, e_{a}{}^M \bar{\nabla}^{\bar{b}} e_{\bar{b} M}  - \bar{\nabla}^{\bar{a}} d \, e_{\bar{a}}{}^M {\nabla}^{b} e_{b M} \right) - \frac{1}{2} \partial_M e_a{}^N \partial_N e^{a M}   \, .
\end{eqnarray}

In order to evaluate the last four terms in ${\cal R}$ it is convenient  to use the form given in 
(\ref{Rsemifinal}). 
The last term in there can be written as 
  \be
 - \frac{1}{4} \Omega_{\bar{c} ab} \Omega^{\bar{c} ab} = - \frac{1}{4} \left(f_{\bar{c} ab} f^{\bar{c} ab} + 2 f_{\bar{c} ab} f^{a b \bar{c}} + 2 f_{a \bar{c} b}f^{a \bar{c} b} + 2 f_{\bar{c} ab} f^{b \bar{c} a} - 2 f_{a \bar{c} b} f^{b \bar{c} a} \right) \, ,
 \ee
 where we have used $f_{A b \bar{c}} =  - f_{A  \bar{c} b}$ which follows from the constraint (\ref{Goff}). 
In total we find 
   \be\label{Interstep314}
   \begin{split}
& -\frac{1}{4} \big(f_{abc} + f_{bca} + f_{cab}\big) f^{cab} - \frac{1}{4} \Omega_{\bar{c} ab} \Omega^{\bar{c} ab}
  \\ 
& \ = \ -\frac{1}{4} \left(f_{C ab} f^{C ab} + 2 f_{C ab} f^{a b C} + 2 f_{a \bar{c} b}f^{a \bar{c} b} + 2 f_{\bar{c} ab} f^{b \bar{c} a} - 2 f_{a \bar{c} b} f^{b \bar{c} a} \right) \, .
 \end{split}
 \ee
The first term in here is zero by the strong constraint. The results for the remaining terms are given by 
  \begin{eqnarray}
    f_{C ab} f^{a b C} &=& - {e_a{}^M} e^{ a L}   \,\partial_N e^{b}{}_{M} \, \partial_L e_b{}^N \;, \\ 
     f_{a \bar{c} b}f^{a \bar{c} b} &=&  \frac{1}{2} \left( e_a{}^M e^{a N} \nabla^b e^{\bar{c}}{}_M \nabla_b 
      e_{\bar{c}N} - e_{\bar{a}}{}^M e^{\bar{a} N} \nabla^{\bar{b}} e^{c}{}_M \nabla_{\bar{b}} e_{c N} \right) \, ,
      \\ 
      f_{\bar{c} ab} f^{b \bar{c} a} - f_{a \bar{c} b} f^{b \bar{c} a} &=&
       - \left( e_{\bar{c}}{}^M e^{\bar{c} N} \nabla_a e_{b M} \nabla^b e^a{}_N - e_{c}{}^M e^{c N} \nabla_{\bar{a}}   
       e_{\bar{b} M} \nabla^{\bar{b}} e^{\bar{a}}{}_N \right)  \\ \nonumber  
       && - e_{\bar{a}}{}^M e^{ \bar{a} L}   \,\partial_N e^{b}{}_{M} \, \partial_L e_b{}^N\;,
 \end{eqnarray}     
which are relatively straightforward to derive by using that the covariant derivatives coincide 
with the ordinary derivatives by arguments similar to those used in (\ref{connectionrel}).  
Inserting these into (\ref{Interstep314}) gives together with (\ref{step1843}) and (\ref{step850}) 
the final result displayed in  (\ref{explR}).

\subsection{Ricci tensor in terms of ${\cal E}_{ij}$}
In this appendix we verify that the Ricci-type tensor ${\cal R}_{a\bar{b}}$ obtained from Siegel's
curvature tensor coincides, upon gauge fixing, with the tensor obtained from the double field theory action (\ref{THEActionINTRO})
by variation with respect to ${\cal E}_{ij}$ as given in \cite{Kwak:2010ew} (there denoted by ${\cal K}_{ab}$).

We start from the expression
\be
 {\cal R}_{a \bar{b}} \ = \ R_{\bar{c} a\bar{b}}{}^{ \bar{c}} \ = \ e_{\bar{c}}\omega_{a\bar{b}}{}^{\bar{c}}-e_{a}
  \omega_{\bar{c}\bar{b}}{}^{\bar{c}}+\omega_{d\bar{b}}{}^{\bar{c}} \omega_{\bar{c}a}{}^{d}
  -\omega_{a\bar{b}}{}^{\bar{d}} \omega_{\bar{c}\bar{d}}{}^{\bar{c}}  \, ,
\ee
which after gauge fixing and upon identification of indices can be rewritten in terms of the 
Christoffel symbols $\Gamma$ reviewed in sec.~\ref{ODDder} as
\be
R_{\bar{k} i\bar{j}}{}^{ \bar{k}} \ = \
{\cal D}_i \Big( \Gamma_{\bar{k} \bar{j}}^{\bar{k}} - \frac{1}{2}  {\cal D}^{k} {\cal E}_{kj} - 2 \bar{\cal D}_j d  \Big) - \bar{\cal D}_{\bar{k}} \Gamma_{i \bar{j}}^{\bar{k}} + \Gamma_{l\bar{j} }^{\bar{k}} \Gamma_{\bar{k}i}^{l} -  \Gamma_{i\bar{j}}^{\bar{l}}  \Big( \Gamma_{\bar{k} \bar{l}}^{\bar{k}} - \frac{1}{2}  {\cal D}^{k} {\cal E}_{kl} - 2 \bar{\cal D}_l d  \Big) \, .
\ee
The dilaton terms reduce to
\be
R_{\bar{k} i\bar{j}}{}^{ \bar{k}} \Big|_d \ = \ 
- 2 {\cal D}_i \bar{\cal D}_j d +  2 \Gamma_{i\bar{j}}^{\bar{l}} \bar{\cal D}_l d
 \ = \ -  ( \nabla_i \bar{\cal D}_j d +  \bar{\nabla}_j {\cal D}_i d) \, ,
\ee
which is consistent with the dilaton terms in ${\cal K}_{i\bar{j}}$ \cite{Kwak:2010ew}, where we note that  
in this appendix all covariant derivatives $\nabla_{i}$ are with respect to $\Gamma$. 

Next, we inspect the $\cal E$-dependent terms in ${\cal R}_{i \bar{j}}$:
\be
R_{\bar{k} i\bar{j}}{}^{ \bar{k}} \Big|_{\cal E} \ = \ {\cal D}_i   \Gamma_{\bar{k} \bar{j}}^{\bar{k}} - \bar{\cal D}_{\bar{k}} \Gamma_{i \bar{j}}^{\bar{k}} + \Gamma_{l\bar{j} }^{\bar{k}} \Gamma_{\bar{k}i}^{l} -  \Gamma_{i\bar{j}}^{\bar{l}}  \Gamma_{\bar{k} \bar{l}}^{\bar{k}} - \frac{1}{2} \nabla_i  {\cal D}^{k} {\cal E}_{kj}  \, ,
\ee
which after some computation can be rewritten as
\begin{eqnarray}
R_{\bar{k} i\bar{j}}{}^{ \bar{k}} \Big|_{\cal E} &=& \frac{1}{2} ( \bar{\nabla}^{k} \bar{\cal D}_k {\cal E}_{ij} 
-  \nabla ^k {\cal D}_i {\cal E}_{kj} -  \bar{\nabla}^k \bar{\cal D}_j {\cal E}_{ik})
  +  \  \frac{1}{4}  g^{pq} ({\cal D}_i {\cal E}_{pj} \bar{\cal D}^k {\cal E}_{qk}  +  {\cal D}_i {\cal E}_{qk} \bar{\cal D}^k {\cal E}_{pj})  \\ \nonumber
  &&  - \frac{1}{4}  (\bar{\cal D}^k {\cal E}_{lj} {\cal D}^l {\cal E}_{ik}  +  \bar{\cal D}^k {\cal E}_{ij} {\cal D}^l {\cal E}_{lk})
    + \ \frac{1}{4}  g^{pq} \bar{\cal D}_j {\cal E}_{ip} {\cal D}^l {\cal E}_{lq}    - \frac{1}{4} g^{kl} g^{pq} \bar{\cal D}_j {\cal E}_{kp} {\cal D}_i {\cal E}_{lq} \, .
\end{eqnarray}
Then $R_{\bar{k} i\bar{j}}{}^{ \bar{k}} = R_{\bar{k} i\bar{j}}{}^{ \bar{k}} \Big|_{\cal E} + R_{\bar{k} i\bar{j}}{}^{ \bar{k}} \Big|_d$ is equivalent to ${\cal K}_{i\bar{j}}$.


\begin{thebibliography}{99}

\bibitem{Giveon:1994fu}
  A.~Giveon, M.~Porrati and E.~Rabinovici,
  ``Target space duality in string theory,''
  Phys.\ Rept.\  {\bf 244}, 77 (1994)
  [arXiv:hep-th/9401139].

\bibitem{Duff:1989tf}
  M.~J.~Duff,
  ``Duality Rotations In String Theory,''
  Nucl.\ Phys.\  B {\bf 335} (1990) 610.

  \bibitem{Tseytlin:1990nb}
A.~A.~Tseytlin,
``Duality Symmetric Formulation Of String World Sheet Dynamics,''
Phys.\ Lett.\ B {\bf 242}, 163 (1990);
``Duality Symmetric Closed String Theory And Interacting Chiral Scalars,''
Nucl.\ Phys.\ B {\bf 350}, 395 (1991).

\bibitem{Hull:2004in}
  C.~M.~Hull,
  ``A geometry for non-geometric string backgrounds,''
  JHEP {\bf 0510} (2005) 065
  [arXiv:hep-th/0406102].

\bibitem{Hull:2006va}
  C.~M.~Hull,
  ``Doubled geometry and T-folds,''
  JHEP {\bf 0707} (2007) 080
  [arXiv:hep-th/0605149].

\bibitem{Hull:2009mi}
  C.~Hull and B.~Zwiebach,
  ``Double Field Theory,''
  JHEP {\bf 0909} (2009) 099
  [arXiv:0904.4664 [hep-th]].


\bibitem{Hull:2009zb}
  C.~Hull and B.~Zwiebach,
  ``The gauge algebra of double field theory and Courant brackets,''
  JHEP {\bf 0909} (2009) 090
  [arXiv:0908.1792 [hep-th]].

\bibitem{Hohm:2010jy}
  O.~Hohm, C.~Hull and B.~Zwiebach,
  ``Background independent action for double field theory,''
  JHEP {\bf 1007} (2010) 016
  [arXiv:1003.5027 [hep-th]].

\bibitem{Hohm:2010pp}
  O.~Hohm, C.~Hull and B.~Zwiebach,
  ``Generalized metric formulation of double field theory,''
  JHEP {\bf 1008} (2010) 008
  [arXiv:1006.4823 [hep-th]].

\bibitem{Berman:2010is}
  D.~S.~Berman and M.~J.~Perry,
  ``Generalized Geometry and M theory,''
  arXiv:1008.1763 [hep-th].

\bibitem{Kwak:2010ew}
  S.~K.~Kwak,
  ``Invariances and Equations of Motion in Double Field Theory,''
  JHEP {\bf 1010} (2010) 047
  [arXiv:1008.2746 [hep-th]].

\bibitem{West:2010ev}
  P.~West,
  ``E11, generalised space-time and IIA string theory,''
  arXiv:1009.2624 [hep-th].

\bibitem{Lust:2010iy}
  D.~Lust,
  ``T-duality and closed string non-commutative (doubled) geometry,''
  arXiv:1010.1361 [hep-th].

\bibitem{Jeon:2010rw}
  I.~Jeon, K.~Lee and J.~H.~Park,
  ``Differential geometry with a projection: Application to double field
  theory,''
  arXiv:1011.1324 [hep-th].

\bibitem{Buscher:1987sk}
  T.~H.~Buscher,
  ``A Symmetry of the String Background Field Equations,''
  Phys.\ Lett.\  B {\bf 194} (1987) 59,
  ``Path Integral Derivation of Quantum Duality in Nonlinear Sigma Models,''
  Phys.\ Lett.\  B {\bf 201} (1988) 466.

\bibitem{Siegel:1993th}
  W.~Siegel,
  ``Superspace duality in low-energy superstrings,''
  Phys.\ Rev.\  D {\bf 48}, 2826 (1993)
  [arXiv:hep-th/9305073].


\bibitem{Siegel:1993xq}
W.~Siegel,
  ``Two vierbein formalism for string inspired axionic gravity,''
  Phys.\ Rev.\  D {\bf 47}, 5453 (1993)
  [arXiv:hep-th/9302036].


 \bibitem{Tcourant}
 T.~Courant, ``Dirac Manifolds."  Trans. Amer. Math. Soc. {\bf 319}: 631-661, 1990.

  %
\bibitem{Hitchin}
N.~Hitchin,
``Generalized Calabi-Yau manifolds,''
 Q. J. Math.  {\bf 54}  (2003), no. 3, 281--308,
arXiv:math.DG/0209099.
%


\bibitem{Gualtieri}
M.~Gualtieri,
``Generalized complex geometry,"
PhD Thesis (2004).
arXiv:math/0401221v1 [math.DG]


\bibitem{Kugo:1992md}
  T.~Kugo and B.~Zwiebach,
  ``Target space duality as a symmetry of string field theory,''
  Prog.\ Theor.\ Phys.\  {\bf 87} (1992) 801
  [arXiv:hep-th/9201040].

\bibitem{Siegel:1999ew}
  W.~Siegel,
  ``Fields,''
  arXiv:hep-th/9912205.





\end{thebibliography}
\end{document}